\documentclass[submission,Phys,letterpaper]{SciPost}
\usepackage{color,amsmath,amssymb,multirow,hyperref,booktabs,graphicx,mathtools} 

\renewenvironment{thebibliography}[1]{
  \begin{oldthebibliography}{#1}
    \setlength{\itemsep}{0em}
    \setlength{\parskip}{0em}
}
{
  \end{oldthebibliography}
}

\newcommand{\arxiv}[1]{\href{http://arxiv.org/abs/#1}{arXiv:#1}}
\newcommand{\urlx}[1]{\href{#1}{#1}}

\newcommand\one{\leavevmode\hbox{\small1\normalsize\kern-.33em1}}

\newcommand{\lag}{\mathcal{L}}

\newcommand{\gev}{\text{GeV}}

\def\slashchar#1{\setbox0=\hbox{$#1$}           
   \dimen0=\wd0                                 
   \setbox1=\hbox{/} \dimen1=\wd1               
   \ifdim\dimen0>\dimen1                        
      \rlap{\hbox to \dimen0{\hfil/\hfil}}      
      #1                                        
   \else                                        
      \rlap{\hbox to \dimen1{\hfil$#1$\hfil}}   
      /                                         
   \fi}

\newcommand{\eg}{\textsl{e.g.}\;}

\newcommand{\be}{\begin{eqnarray*}}
\newcommand{\ee}{\end{eqnarray*}}

\newcommand{\bee}{\begin{eqnarray}}
\newcommand{\eee}{\end{eqnarray}}
\newcommand{\beeq}{\begin{equation}}
\newcommand{\eeeq}{\end{equation}}

\begin{document}

\begin{center}{\Large \textbf{
CapsNets Continuing the Convolutional Quest
}}\end{center}

\begin{center}
S.~Diefenbacher\textsuperscript{1},
H.~Frost\textsuperscript{2},
G.~Kasieczka\textsuperscript{2},
T.~Plehn\textsuperscript{1}, and
J.~M.~Thompson\textsuperscript{1}
\end{center}

\begin{center}
{\bf 1} Institut f\"ur Theoretische Physik, Universit\"at Heidelberg, Germany \\
{\bf 2} Institut f\"ur Experimentalphysik, Universit\"at Hamburg, Germany\\

plehn@uni-heidelberg.de
\end{center}

\begin{center}
\today
\end{center}

\section*{Abstract}
{\bf Capsule networks are ideal tools to combine event-level and
  subjet information at the LHC.  After benchmarking our capsule
  network against standard convolutional networks, we show how
  multi-class capsules extract a resonance decaying to top quarks from
  both, QCD di-jet and the top continuum backgrounds. We then show how
  its results can be easily interpreted. Finally, we use associated
  top-Higgs production to demonstrate that capsule networks can work
  on overlaying images to go beyond calorimeter information.}

\vspace{10pt}
\noindent\rule{\textwidth}{1pt}
\tableofcontents\thispagestyle{fancy}
\noindent\rule{\textwidth}{1pt}
\vspace{10pt}

\clearpage
\section{Introduction}
\label{sec:intro}

New developments in machine learning have recently started to
transform different aspects of LHC physics. The most visible
development is, arguably, deep learning in subjet physics. The
underlying idea is to replace multi-variate analyses of high-level
observables by deep neural networks with low-level observables. It
follows directly from our improved understanding of subjet
physics both experimentally and theoretically, and from the rapid
development of standard machine learning tools~\cite{early_stuff}.

A standard approach to deep learning of jet physics is based on jet
images, where we extract information from heat maps in the rapidity vs
azimuthal angle plane~\cite{jet_images,aussies}. Several studies have
looked at what information a neural network can extract from
jets~\cite{information}. The most relevant measurements come from the
calorimeter and need to be combined with tracking
information. Standard benchmarks based on jet images or alternative
network setups are quark-gluon discrimination~\cite{jets_qg},
$W$-tagging~\cite{jets_w}, Higgs tagging~\cite{jets_h}, and
top-tagging~\cite{jets_top,deep_top1,deep_top2,lola}. This relatively
straightforward classification task only served as a first attempt of
deep learning in LHC analyses~\cite{jets_comparison}, and the progress
in this field should encourage us to search for more challenging and
transformative applications. One promising line of reserach is related
to ways of training neural networks at the LHC, for instance using
weakly supervised learning~\cite{weak}, unsupervised
classification~\cite{unsupervised}, or unsupervised
autoencoders~\cite{auto}. Alternatively, we can extend our
classification task trained on data to include statistical and
theoretical uncertainties~\cite{bayes}.\medskip

Once we have sufficient control over the network training we can come
back to modern LHC physics, where jets have turned from the main
objects of event analyses to a somewhat arbitrary separation line
between subjet analysis and event-level analyses. The question becomes
how neural network architectures cope with the full event
information. We emphasize that such event information should again be
low-level observables rather than a small number of 4-vectors
describing the hard process at this high level~\cite{simple_nn}.  A
natural extension of convolutional networks on event and jet
images~\cite{events_cnn,end_to_end} are capsule
networks~\cite{capsules1,capsules2}. For applications in
astrophysics, see \eg Ref.~\cite{capsules_astro}. The main advantage
of capsules is in analyzing structures of objects and simultaneously
their geometric layout. It perfectly matches our task of combining
subjet information with the event-level kinematics of jets and other
particles.\medskip

In this paper we start with a brief introduction to capsule networks
as an extension of convolutional networks in
Sec.~\ref{sec:capsules}. Next, we apply capsule networks to the
classification of di-top events at the subjet level in
Sec.~\ref{sec:tagging}. This allows us to benchmark our capsule
network with established machine learning top taggers using for
example convolutional networks~\cite{deep_top1,deep_top2}. Next, we
separate full events corresponding to a $Z' (\to t\bar{t})$ signal
from $t\bar{t}$ and from di-jet backgrounds in
Sec.~\ref{sec:reco}. Here we introduce multi-class capsules to control
the different backgrounds. In Sec.~\ref{sec:inside} we how these
results can be visualized especially well.  Finally, in
Sec.~\ref{sec:higgs} we consider a challenging application, the
semi-leptonic final state of $t\bar{t}H_{bb}$ production. It allows us
to explore the full power of capsule networks to extract information
from overlaying images, going beyond calorimeter images and opening a
path towards tracking information~\cite{seeing}.

We emphasize that in this study we do not discuss the issues related
to training networks for event tagging or the systematic and
theoretical uncertainties related to it. They need to be tackled
now that we know how large event images can be efficiently analyzed by
capsule networks --- as the natural extension of convolutional
networks working on jet images.

\section{Capsule networks}
\label{sec:capsules}

In this paper we introduce capsule neural networks (CapsNets) as a
natural replacement for standard convolutional neural networks in LHC
physics.  We refer to the established convolutional networks as scalar
CNNs because they rely on single numbers. CapsNets replaces these
single numbers with capsule vectors describing the feature maps.  For
example, 24 feature maps with ${40 \times 40}$ entries each could be
defined as 1600 capsules of dimension 24, or 3200 capsules of
dimension 12, or 4800 capsules of dimension 8, etc. Each capsule can
be thought of as a vector in signal or background feature space,
depending on which it describes. The length of this vector then
encodes how signal-like or background-like the image is. The idea
behind these vectors is that they can track the actual geometric
position and orientation of objects, which is useful for images
containing multiple different objects. In particle physics, an entire
event image is a perfect example of this.

Just like a scalar CNN, a CapsNet starts with a pixelled image, in
our case the calorimeter image of a complete event with $180 \times
180$ pixels. This image is analyzed with a convolutional filter, for
example extracting edges. The size of these kernels is not fixed, so
one way of reducing the size of a sparsely filled image is to choose
kernels with at least ${(n+1) \times (n+1)}$ pixels and to move $n$
rows or columns per step. This is known as a convolution with stride
$n$, in contrast to pooling layers which simply decrease the
resolution of the image. How significant the difference is between
these approaches depends on the details of the
analysis~\cite{capsules1,capsules2,striving}. Our CapsNets include
several layers of convolution with multiple feature maps. They extract
the relevant information from the input image, and is so far identical
to a scalar CNN. The advantages and power of the CapsNet come from the
additional capsule layers after the convolutions.\medskip

\begin{figure}[t]
\centering
\includegraphics[width=0.7\textwidth]{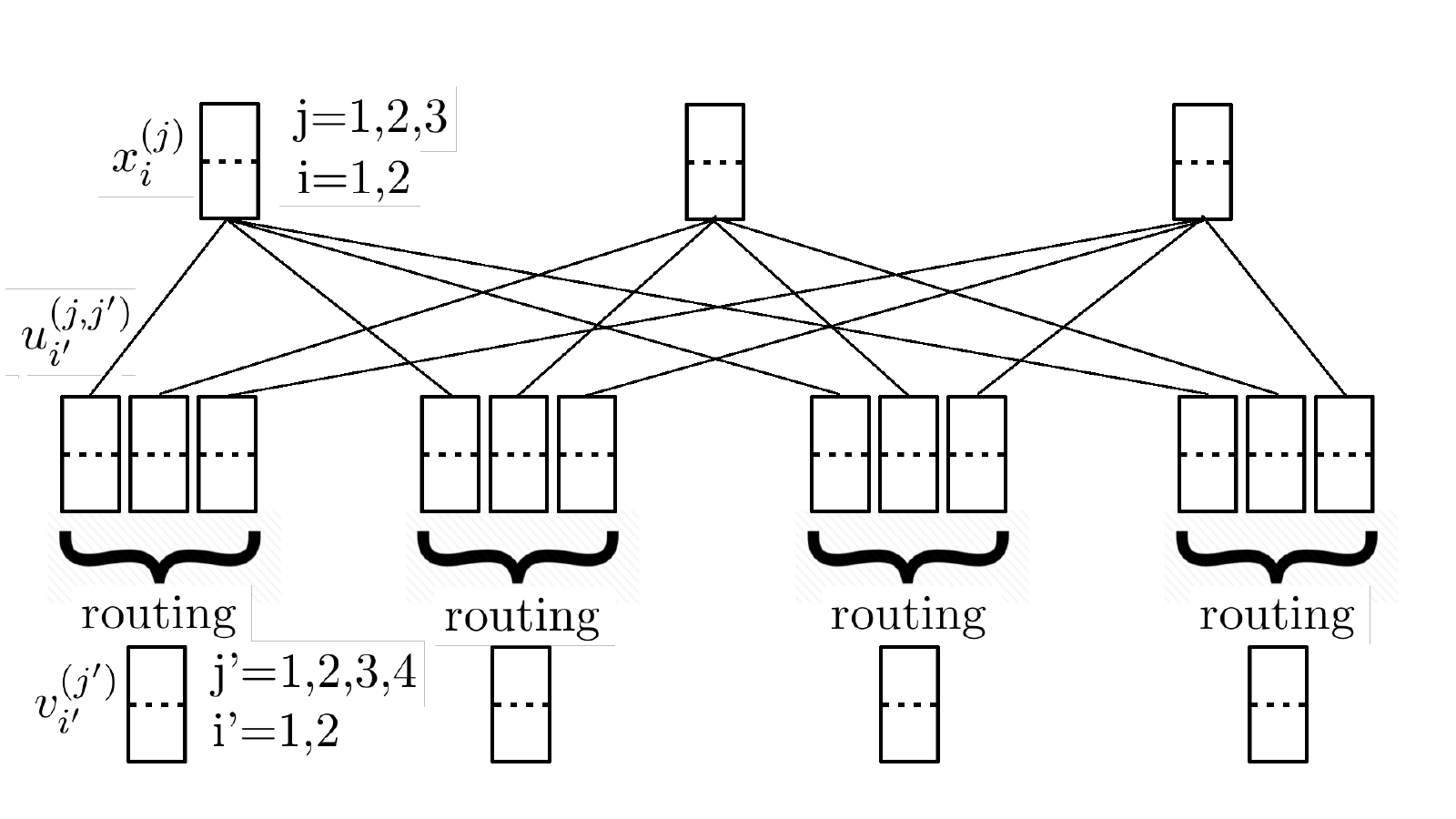}
\caption{Sketch of a CapsNet module with two simple capsule layers.}
\label{fig:Capsnet}
\end{figure}

Deep CapsNets consist of several capsule layers.  After the
convolution part of the CapsNet, each layer consists of a number of
parallel capsules. These capsules have to transfer information
matching their vector property.  In Fig.~\ref{fig:Capsnet} we
illustrate a small, two-layer CapsNet with three initial capsules
$\vec{x}^{(j)}$ of dimension two linked through routing by
agreement~\cite{capsules1} to four capsules, also of dimension two,
\begin{align}
x_i^{(j)} \longrightarrow v^{(j')}_{i'} 
\quad \text{with} \quad i=1,2 \quad i'=1,2 \quad j=1,2,3 \quad j'=1,2,3,4 \; .
\end{align}
For deeper networks the dimensionality of the resulting capsule vector
can, and should, be larger than the incoming capsule vector. However,
in our illustration we keep the dimensionality of the
capsules at two for clarity.  To get from three to four
capsules we first define four combinations of the three initial
capsules. Their entries are defined as $u^{(j,j')}_{i'}$, and they are
related to the initial capsule vectors $\vec{x}^{(j)}$ through
trainable weights,
\begin{align}
u^{(j,j')}_{i'} = \sum_{i=1,2} w^{(j,j')}_{i' i} \; x^{(j)}_i \; ,
\label{eq:weight_u}
\end{align}
as indicated by the arrows in Fig.~\ref{fig:Capsnet}. The assignment
of lower and upper indices in our description only serves
illustrational purposes. Next, we need to contract the index $j$ to
define the four outgoing capsules. For this purpose we define another
set of trainable weights and write
\begin{align}
v_{i'}^{(j')} = \sum_{j=1,2,3} c^{(j,j')} \; u^{(j,j')}_{i'} \; .
\label{eq:output_capsule}
\end{align}
These weights $c^{(j,k)}$ get normalized through a SoftMax operation
\begin{align} 
\sum_{j'=1,2,3,4} c^{(j,j')} &= 1 \quad \forall j \notag \\
c^{(j,j')} &= \text{SoftMax}_{j'} \; c'^{(j,j')} 
= \frac{\exp c'^{(j,j')}}{\sum_\ell \exp c'^{(j,\ell)}}\, ,
\label{eq:weight_norm}
\end{align}
on a set of general weights $c'$.  This ensures that the contributions from one capsule in the former to each capsule in the current layer add up to one.
Furthermore, the squashing step applied after
each capsule layer ensures that the length of each output
capsule vector remains between 0 and 1,
\begin{align}\label{eq:squash}
v_{i'}^{(j')} = \vec v \to \vec v' &= \frac{\vec v^2}{1+ \vec v^2} \; \hat{v}\, \\
\vec v \to \vec v' &= \frac{|\vec v|}{\sqrt{1+ |\vec v|^2}} \; \hat{v}\,,
\label{eq:squash2}
\end{align}
with $\hat{v}$ defined as the unit vector in $\vec v$-direction.  The
advantage of Eq.\eqref{eq:squash2} over Eq.\eqref{eq:squash} is that
it does not iteratively shrink small inputs to zero. This feature
becomes important when considering multiple capsule layers.\medskip

\begin{figure}[t]	
\centering
\includegraphics[width=0.2\textwidth]{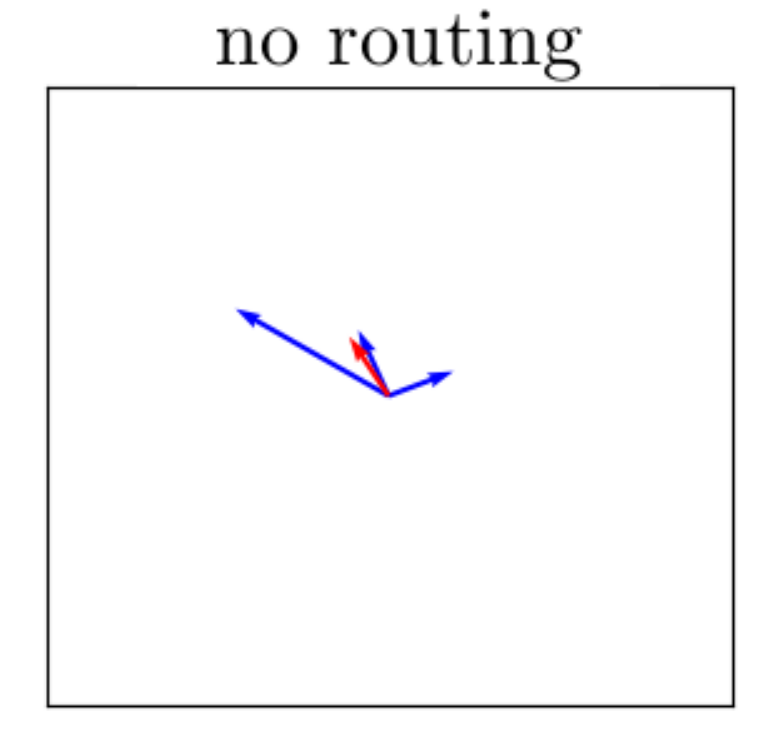}
\includegraphics[width=0.2\textwidth]{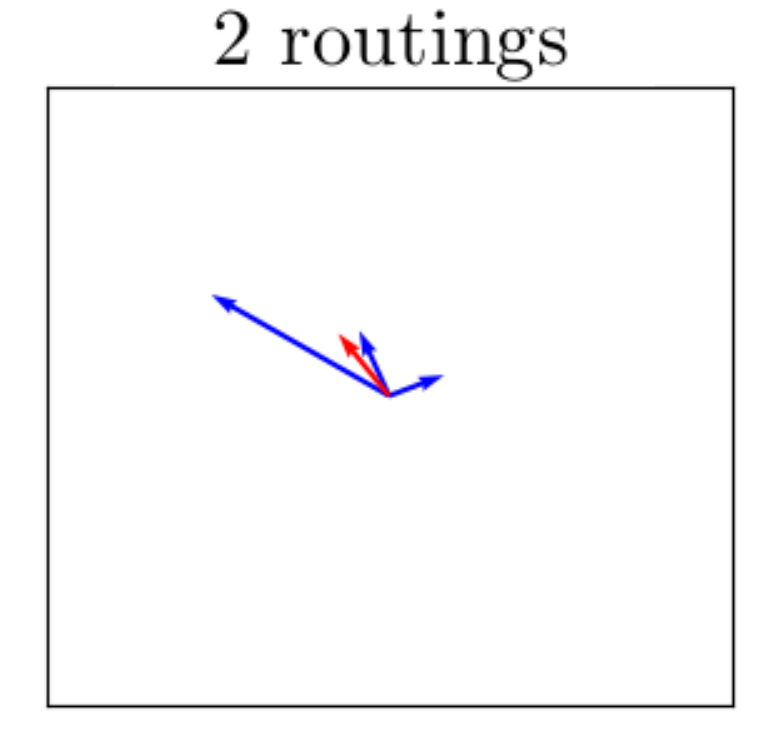}
\includegraphics[width=0.2\textwidth]{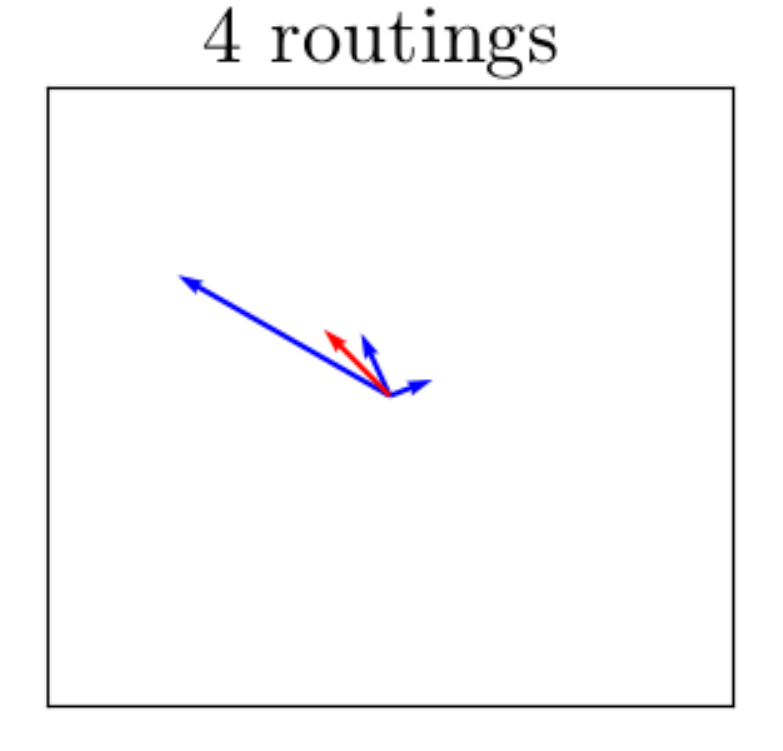}
\includegraphics[width=0.2\textwidth]{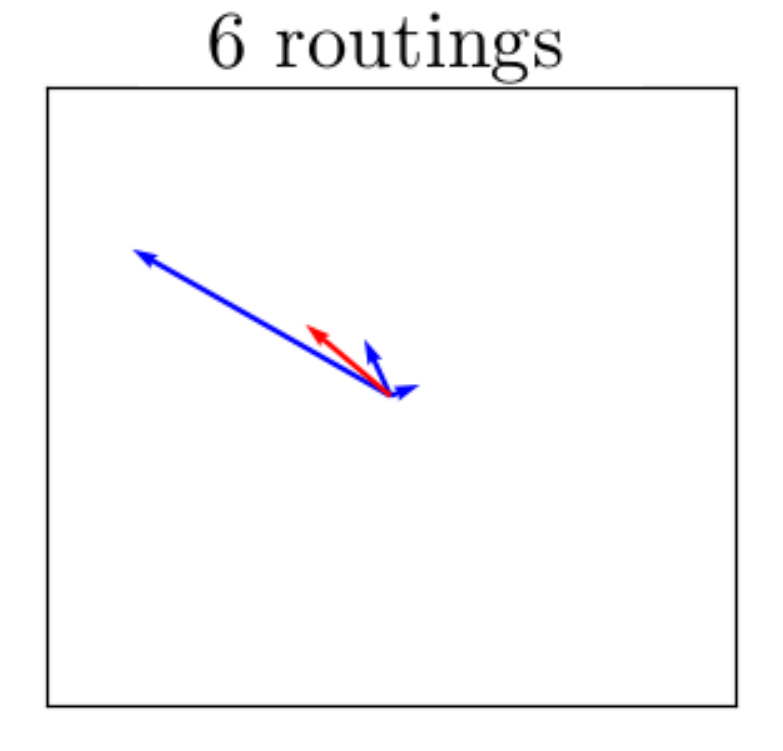}
\caption{Effects of the routing/squashing combination. In blue we show
  the intermediate vectors, in red we show the output vector after
  squashing.}
\label{fig:routing_illus}
\end{figure}

Up to now we have constructed a set of four capsules from a
set of three capsules through a number of trainable weights, but not
enforced any kind of connection between the two sets of capsule
vectors.  We can extend the condition of Eq.\eqref{eq:weight_norm} to
consecutively align the vectors $\vec{u}^{(j,j')}$ and $\vec{v}^{(j')}$
through a re-definition of the weights $c^{(j,j')}$. This means we
compute the scalar product between the vector $\vec{u}^{(j,j')}$ and
the squashed vector $\vec{v}^{(j')}$ and replace
\begin{align}
c'^{(j,j')} \longrightarrow c'^{(j,j')} + \vec{u}^{(j,j')} \cdot \vec{v}^{(j')} \; ,
\end{align}
which converges once $\vec{u}^{(j,j')}$ and $\vec{v}^{(j')}$ are
parallel. We apply this replacement to each capsule, or fixed $j'$,
individually, before we once again apply the SoftMax operation. We
repeated this for 3 routings, which has been shown in other studies to
give the best results~\cite{capsules1}. The
routing is illustrated in Fig.~\ref{fig:routing_illus}, where the blue
vectors represent the three intermediate $\vec{u}^{(j,j')}$ in each
set and the red vector is the combination $\vec{v}^{(j')}$. We can see
how, with each routing iteration, the vectors parallel to
$\vec{v}^{(j')}$ become longer while the others get shorter.\medskip

In the CapsNet framework we use the squashed length
of the output vectors $\vec{v}^{(j')}$ for classification. In complete
analogy to the scalar CNN we differentiate between signal and
background images using two output capsules. The more likely the image
is to be signal or background, the longer the output capsule vectors
will be. The corresponding margin loss for a set of output capsules $j'$ is
defined as 
\begin{align}
L 
&= \sum_{j'} L^{(j')}  \notag \\
L^{(j')} 
&= T^{(j')} \max \left( 0,m_+ - | \vec{v}^{(j')} | \right)^2 
+ \lambda (1 - T^{(j')}) \max \left( 0, | \vec{v}^{(j')} |- m_- \right)^2 \; .
\label{eq:capsoutput}
\end{align}
$T^{(j')}$ is the truth label of the input, so for a simple
classification task we use $T^{(1)} = 1$ and $T^{(2)} = 0$ and the
loss function consists of the two terms
\begin{align}
L^{(1)} &= \max \left( 0,m_+ - | \vec{v}^{(1)} | \right)^2  \notag \\
L^{(2)} &= \lambda \max \left( 0, | \vec{v}^{(2)} |- m_- \right)^2 \; .
\end{align}
Using, for example, $m_+ = 0.9$ means that the network will seek
signal vectors $\vec v^{(1)}$ with length above $0.9$, where the loss
vanishes. Similarly, for $m_-=0.1$ the network prefers background
vectors $\vec v^{(2)}$ shorter than 0.1.  While these target numbers
of the capsule length, $0.9$ and $0.1$, sum up to unity, nothing
forces the actual length of all capsules in a prediction to do the
same. Using the $\lambda$ parameter we can scale the importance of the two terms in the loss function. We chose $\lambda = 0.5$, putting the main emphasis on having the correct capsule length being close to the target number. Although not relevant for the conclusions in this study, the CapsNet
receives an additional term in its loss function from the
reconstruction of the initial image from the outputs.

For a 2-class classification task it should be possible to define a
simpler setup where one output capsule encodes the entire signal vs
background information.  This also means that for our setup the
capsule length output $|\vec v^{(i)}|$ cannot be linked to a
probability, but as a set of scores which describe the how signal-like
or background-like an event is. Combining them into a single
classifier is not unique, as we will see later. We will also see that
the LHC analyses we propose in this paper are not 2-class
classification tasks, so we keep our multi-capsule output for
now.\medskip

The advantage of the CapsNet over the scalar CNN is that each entry of
the capsule vector can learn certain features independently of the
other entries, and only the combination of all entries is required to
separate signal and background. This flexibility should, for instance,
replace the pre-processing which is often used for jet
images~\cite{deep_top1,deep_top2}. Following the original test cases
for capsule networks, the vector entries can also learn individual
patterns independently from the geometric arrangement. This is
precisely what we will exploit in our combination of subjet and
event-level patterns in LHC events.  We implement all our CapsNets
with the \textsc{Keras} \textsc{Python} package~\cite{keras} and a
\textsc{TensorFlow}~\cite{tensorflow} back-end. We also make use of
the usual \textsc{Adam} optimizer~\cite{adam}.

\section{Di-top tagging}
\label{sec:tagging}

Before we use CapsNets to analyze full events, we need to
confirm that they successfully analyze subjet structures. As an
experimentally and theoretically safe, established benchmark we use
top tagging~\cite{top_tagger,heptop1,heptop2,jets_comparison}, specifically the signal
process
\begin{align}
pp \to Z' \to t \bar{t}
\label{eq:signal}
\end{align}
for $m_{Z'}= 1$~TeV and a narrow width of $\Gamma_{Z^\prime}=1$~GeV.
A small width is useful when we eventually extract the narrow
resonance from a continuum background.  Because this is a BSM process,
we first generated the model with
\textsc{FeynRules}~\cite{feynrules}.  This simplified $Z'$ model
extends the Standard Model Lagrangian by
\begin{align}
\lag_{Z'} = c_1 \sum_{q} c_q \bar{q} \gamma_\mu q Z^{\prime \mu} \; ,
\end{align}
where $c_1$ and $c_q$ are freely chosen constants determining the
normalization of the signal. The $Z^\prime$ decays to a $t\bar{t}$
pair, which in turn decay hadronically.\medskip

Such a $t\bar{t}$ resonance search allows us to split the analysis
into two steps~\cite{zprimetoptag}.  First, we focus on the
subjet-level information from the two fat jets and ignore their
event-level kinematics. For this purpose we limit ourselves to the
light-flavor di-jet background
\begin{align}
pp \to jj 
\qquad \text{with} \; j= g, u, d, c, s, b\; ,
\label{eq:qcd_background}
\end{align}
and will add the continuum top pair background in the next
section.\medskip

All events in this study are for a 13~TeV LHC.  We generate
leading-order events with \textsc{Sherpa}2.2.5~\cite{sherpa} and the
\textsc{Comix} matrix element generator~\cite{comix},
where we enable parton shower and hadronization effects. We neglect
underlying event and pile-up, which we assume can be removed by
dedicated tools~\cite{pileup}. We use the 5-flavor LO
\textsc{NNPDF}3.0 PDF set~\cite{nnpdf}.  Detector effects are
simulated with \textsc{Delphes}~\cite{delphes} and a standard ATLAS
card with a modified jet radius and jet algorithm for each
process. All jets are defined by \textsc{FastJet}~\cite{fastjet}.  In
this section jets are defined by the C/A jet clustering
algorithm~\cite{CA}, with $R=1.0$ and
\begin{align}
p_{T,j} > 350~\gev 
\qquad \text{and} \qquad 
|\eta_j| < 2.0 \; .
\end{align}

\begin{figure}[t]
\includegraphics[width=0.99\textwidth]{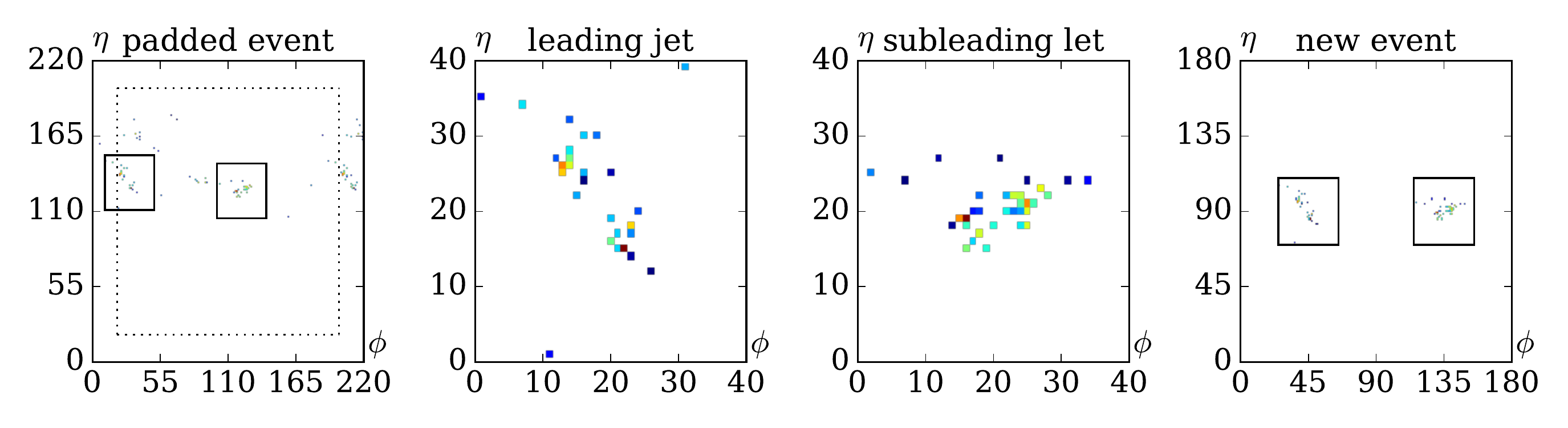}
\caption{Processing of the event images to a pair of top images.}
\label{fig:ditop_visualization}
\end{figure}

From this output we extract the calorimeter hits and transform them
into a 2D jet image with $E_T$ as the pixel value. The calorimeter images have a
size of
\begin{align}
180 \times 180~\text{pixels,}
\end{align}
covering $|\eta| < 2.5$ and ${\phi = 0~...~2\pi}$. The periodicity in
$\phi$ is accounted for by phi-padding with an appropriate depth or
number of repeated pixels in both positive and negative $\phi$
direction. We separately choose the amount of padding for each
convolutional layer and equal to half the respective kernel size. For
this benchmarking exercise we then remove all event-level information,
such as $\eta$ and $\phi$ positions of the jets. We take the
event-level calorimeter images, pad them with zeros in $\eta$ and
symmetrically in $\phi$ and select $40 \times 40$ pixel sections
around the axes of the two leading jets. Two such jet images are then
pasted back into empty $180\times 180$ images.  This process is
illustrated in Fig.~\ref{fig:ditop_visualization}.\medskip

\begin{figure}[t]
\begin{center}
\includegraphics[width=0.8\textwidth]{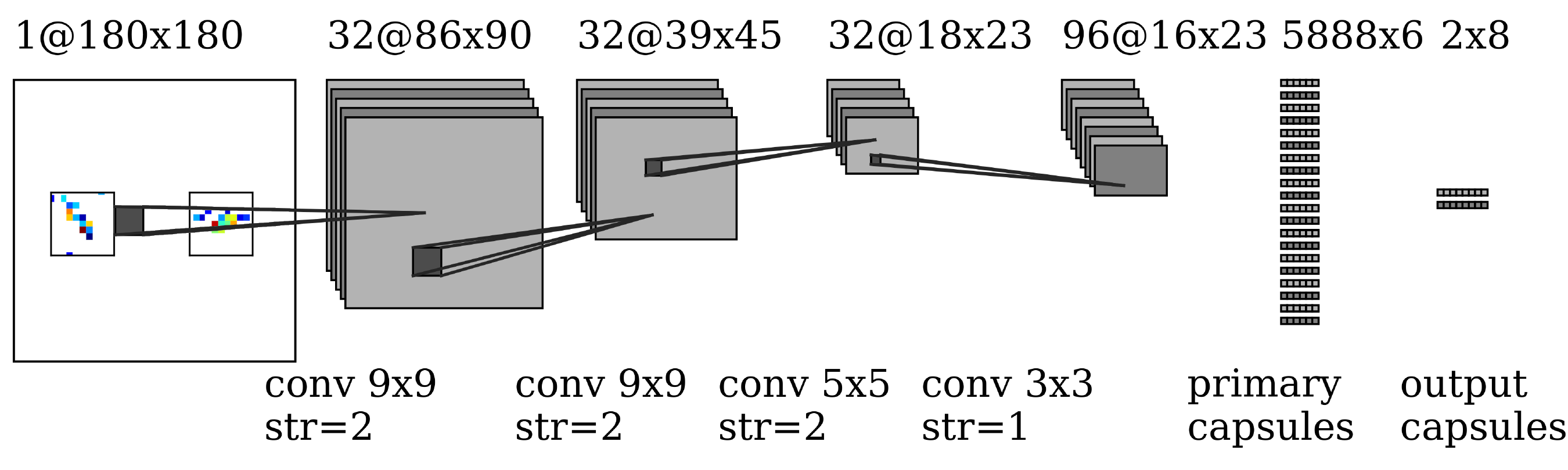} \\[10mm]
\includegraphics[width=0.8\textwidth]{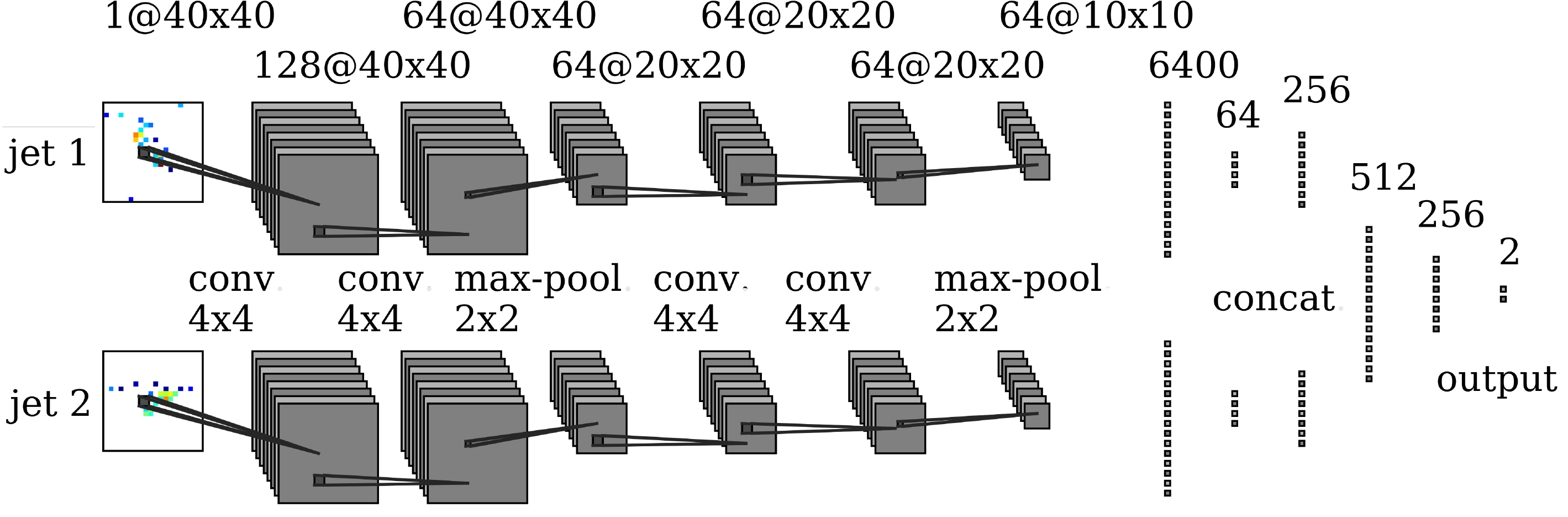} \\[10mm]
\includegraphics[width=0.8\textwidth]{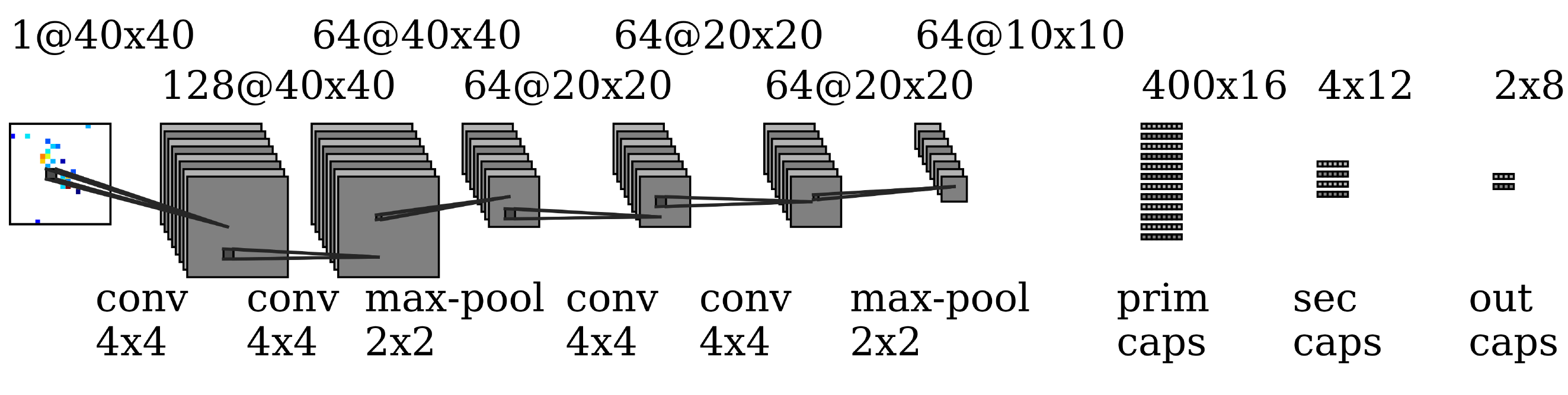}
\end{center}
\caption{Top: convolutional CapsNet architecture based on
  stride-convolutions, used for di-Top tagging.  Center: scalar CNN
  di-top tagger using the Rutgers \textsc{DeepTop}
  architecture~\cite{deep_top1,deep_top2}.  Bottom: CapsNet single top
  jet tagger architecture.}
\label{fig:capsnet_conv}
\end{figure}

These simplified (${180 \times 180}$)-pixel event images are what we
feed into our CapsNet shown in the top panel of
Fig.~\ref{fig:capsnet_conv}. Our architecture avoids pooling and
instead uses convolutions with stride two, as outlined in
Sec.~\ref{sec:capsules}. We produce 32 feature maps and for the first
two layers we use a $9\times 9$ kernel and a stride of two. Then we
reduce the kernel size to $5\times 5$ for the third layer, still with
a stride of two. Finally, we apply one regular convolution with a
stride of one and a $3\times 3$ kernel.  With this final convolution
we also increase the number of feature maps to 96.

Analogous to the original capsule paper~\cite{capsules2}, we transition 
between the convolutional and capsule parts by re-shaping the output of the
convolutional layers into a capsule layer with $j \le 5888$ capsules
of dimension $i \le 6$ and add a second layer with dimension $i' \le
8$ and $j'=1,2$ capsules, which are used as outputs for the
classification. Here $i^{(')}$ and $j^{(')}$ run over the
dimensionality and number of capsules, respectively, as described in
Sec.~\ref{sec:capsules}.\medskip

If we want to evaluate the performance of our network we need an
estimator to build an ROC curve.  In a scalar CNN with SoftMax
activation in the final layer we usually use the output of the signal
neuron, because the background neuron does not give an independent
result. The equivalent approach for our CapsNet could rely on the
length of the signal capsule, $|\vec{v}^{(s)}|$. However, our CapsNet
does encode additional information in the background capsule, so based
on the output capsules in Eq.\eqref{eq:capsoutput} we can define
estimators of the kind
\begin{align}
|\vec{v}^{(S)}| 
\qquad \text{or} \qquad
|\vec{v}^{(S)}| 
- |\vec{v}^{(B,1)}| 
- |\vec{v}^{(B,2)}| 
- \cdots
\label{eq:signalness}
\end{align}
This choice affect the tagging performance for realistic
training. Throughout the paper we will use the second choice as the
default.

\begin{figure}[t]
\centering
\includegraphics[width=0.45\textwidth]{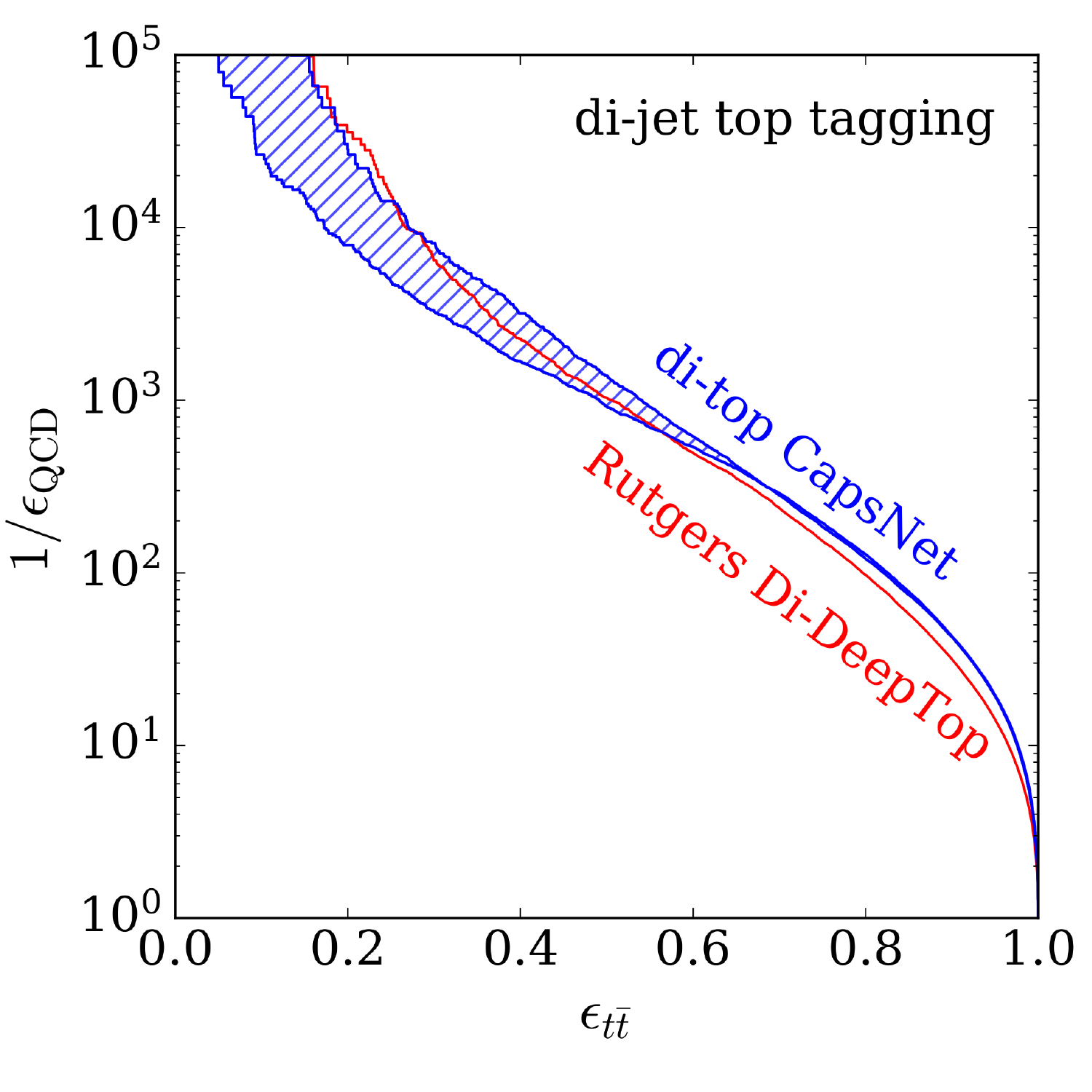}
\hspace*{0.03\textwidth}
\includegraphics[width=0.45\textwidth]{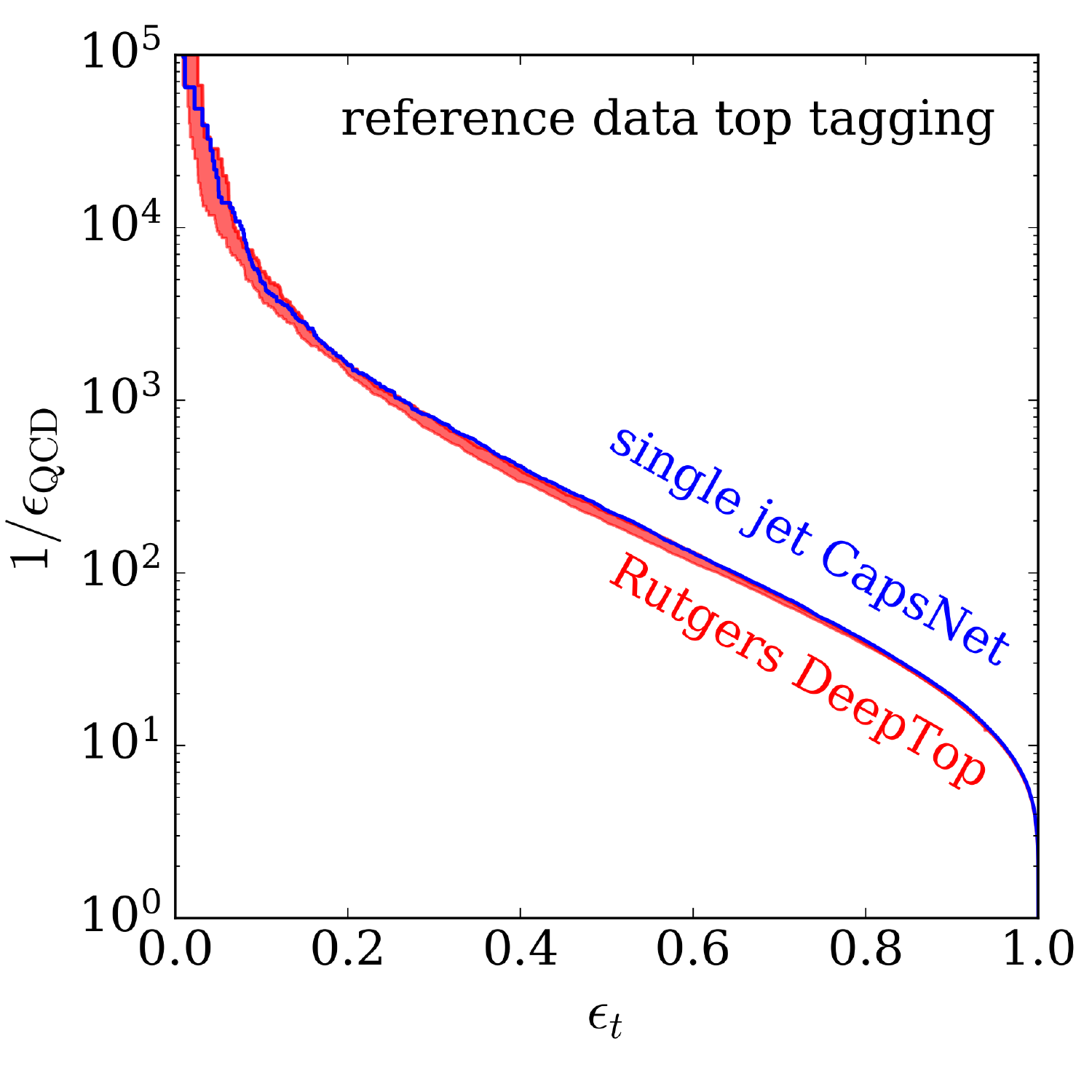}
\caption{Left: comparison between CapsNet and the Rutgers
  \textsc{DeepTop} CNN for for the di-tops vs QCD di-jets. Right: same
  comparison for single top jets using a standard
  dataset~\cite{dataset,jets_comparison}.}
  \label{fig:vs_DeepTop_roc_curve}
\end{figure}

We can now compare the performance of our CapsNet to a combination of
two scalar CNN taggers, specifically the Rutgers \textsc{DeepTop}
tagger~\cite{deep_top1,deep_top2}. In contrast to the minimal pre-processing
we use for the event image capsule network, for the Rutgers tagger and the 
jet image capsule network we employ 
the full pre-processing for each jet as described in Ref.~\cite{deep_top2}.
The jets are selected and centered around the $p_T$ weighted centroid of the
jet, and rotated such that the major principal axis is vertical.
The image is then flipped to ensure that the maximum activity is in
the upper-right-hand quadrant. Finally, the images are pixelated and
normalized.

It is shown in the center panel of
Fig.~\ref{fig:capsnet_conv}. We use a total of 500,000 events, split
into three parts training and one part each for testing an
validation. For training we use the \textsc{Adam} optimizer with a
learning rate of 0.001 and a decay of 0.9, and we employ early
stopping to interrupt training once the validation accuracy stops
increasing. The result of this comparison can be seen in the left
panel of Fig.~\ref{fig:vs_DeepTop_roc_curve}. The shaded curve
represents the two estimators given in Eq.\eqref{eq:signalness}, where
in this case the signal capsule alone gives the better results and an
ROC curve more compatible with the 2-class scalar CNN.  Within this
uncertainty, the CapsNet performs as well as two copies of a dedicated
tagger for the subjet information alone. Given that the CNN is
well-optimized, this is the best we can expect for our relatively
straightforward CapsNet.\medskip

To allow for a direct comparison with many other tools, we also apply
our CapsNet to single top jets from a public
dataset~\cite{dataset,jets_comparison}, based on events generated for
the study in Ref.~\cite{lola}. In that case there exists no
event-level information for the CapsNet shown in the bottom panel of
Fig.~\ref{fig:capsnet_conv}. Again, the CapsNet turns out competitive
with state-of-the-art convolutional networks, albeit not quite with
the leading tools presented in Ref.~\cite{jets_comparison}.

\section{Di-top resonance}
\label{sec:reco}

\begin{figure}[t]
\centering
\includegraphics[width=0.9\textwidth]{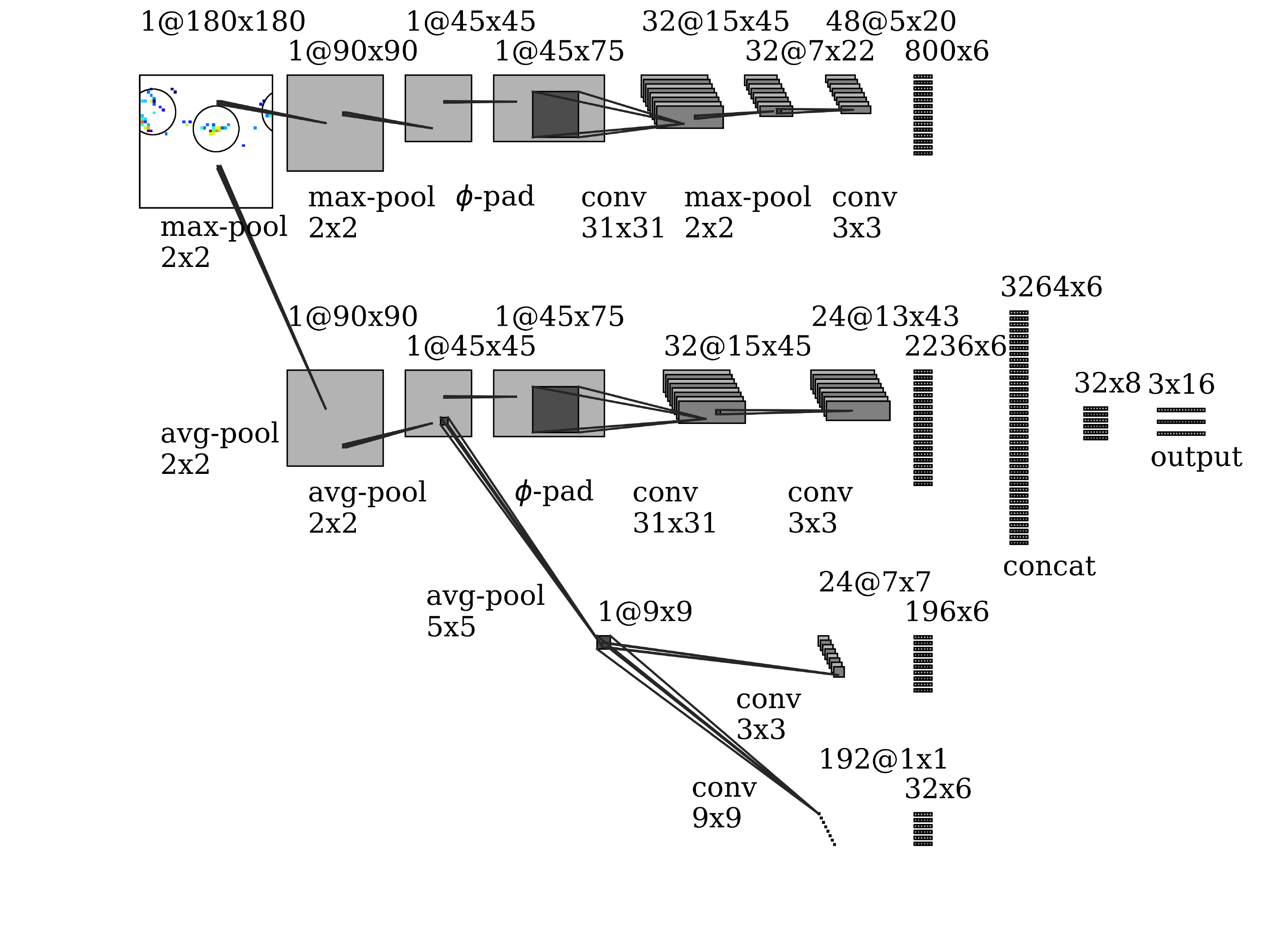}
\caption{Pooling CapsNet architecture. The re-shaping of the final
  convolution results into capsules with dimension $i \le 6$ is not
  detailed.}
\label{fig:pooling_capsnet}
\end{figure}

In our second benchmarking step we now consider event-level
information.  To see how the CapsNet uses event kinematics in addition
to subjet-level information we first study the continuum $t\bar{t}$
background to our $Z'(\to t\bar{t})$ signal,
\begin{align}
pp \to t\bar{t} \qquad \text{(SM).}
\label{eq:top_background}
\end{align}
For this classification the subjet information does not help.  Only
the combination of the continuum $t\bar{t}$ and QCD backgrounds then
requires the CapsNet to learn both the geometry of the event and the
subjet differences of top and QCD jets.\medskip

Now that the signal jets and background jets are both boosted top
quarks, the network needs to rely on the $Z'$ kinematics and
differences in radiation patterns between signal and background.  On
the CapsNet side the increased complexity of the full events leads to
a slightly more involved architecture than the one described in
Sec.~\ref{sec:tagging}.  Our new architecture combines max-poolings,
average-poolings, and convolutions, to make it easier to (also) focus
on large-scale features. It is shown in
Fig.~\ref{fig:pooling_capsnet}. The idea behind the setup is that (i)
the max-pooling preserves the highest value pixels, allowing the
network to learn both the absolute and relative jet positions, and
(ii) the average-pooling preserves the total transverse
energy. Consequently, we use two different resolutions, $45 \times 45$
to learn the jet $E_T$ and $9\times 9$ to learn the total energy in
the event. The two different pooling strategies are implemented as two
parallel branches in the network. The average pooling branch is
further subdivided into three branches. This allows for three
different kernel sizes to be used in parallel, corresponding to three
different scales of activity.

As the benchmark for the event-level analysis we use a boosted
decision tree with the \textsc{SciKit-learn} \textsc{Python}
package~\cite{scikit} and using \textsc{AdaBoost}~\cite{adaboost}.  No
cuts are placed on the $t\bar{t}$ final state, and the events are
generated according to the procedure described in
Sec.~\ref{sec:tagging}.  We give the BDT the event-level information
\begin{align}
\{m_{jj},\,p_{T j_1},\, p_{T j_2},\, \eta_{j_1}, \,\eta_{j_2}\}\,.
\end{align}
The BDT has a maximal depth of three and uses 100 estimators. Training
and testing of the BDT are performed with the same samples used for
the network training and evaluation. We use 500,000 events, split into
300,000 training events, 100,000 testing events and 100,000 validation
events.

\begin{figure}[t]
\centering
\includegraphics[width=0.45\textwidth]{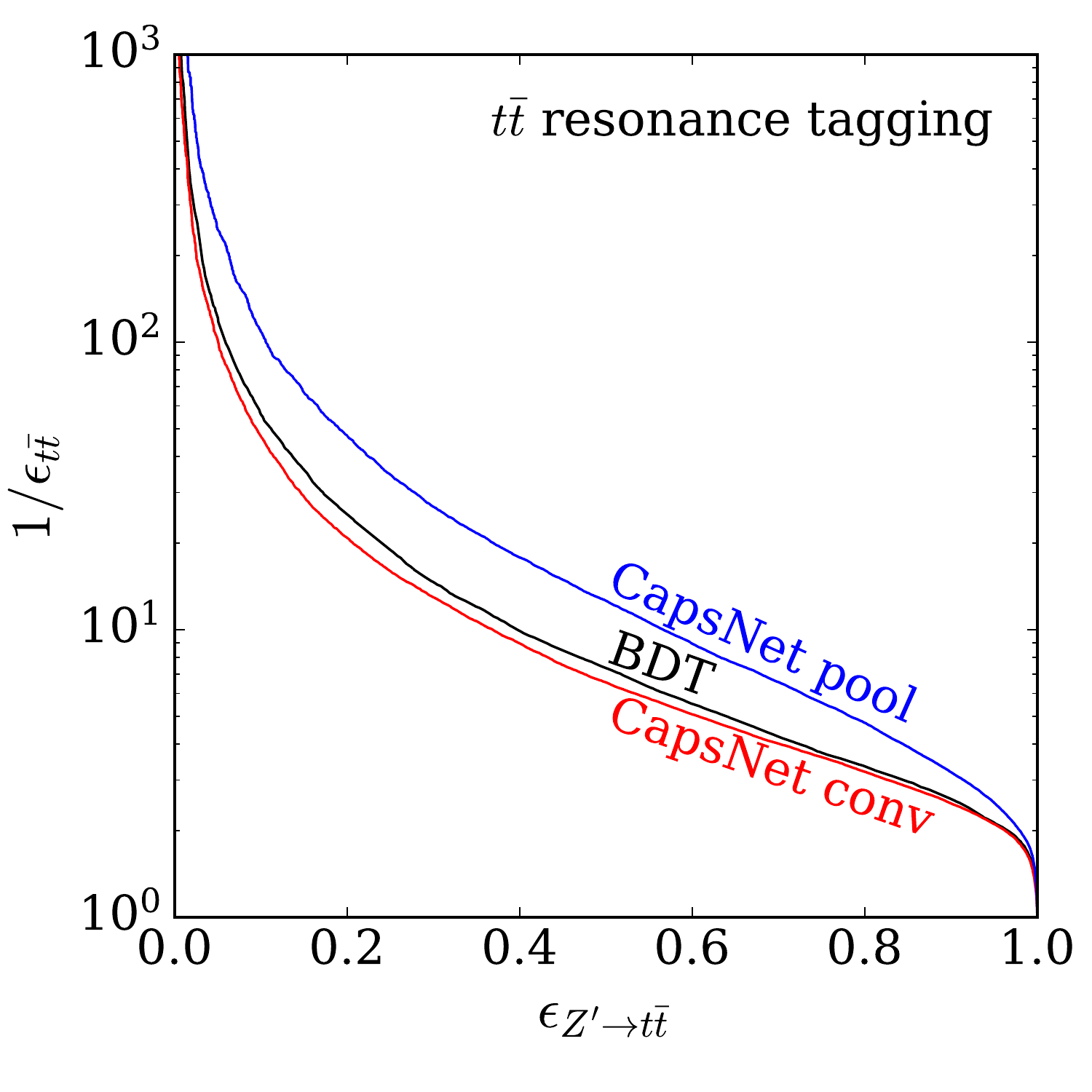}
\caption{ROC curves for two capsule networks and the BDT benchmark,
  trained and tested on $Z^\prime(\to t\bar{t})$ signal and continuum
  $t\bar{t}$ background events.}
\label{fig:vsBDT_roc}
\end{figure}

Figure~\ref{fig:vsBDT_roc} shows that our extended CapsNet
architecture performs significantly better than both the simpler
CapsNet and the BDT baseline.  Specifically, the convolutional CapsNet
slightly under-performs the BDT, whereas the larger architecture is
more able to describe the complexity of a full event, leading to a
significant improvement over a simple BDT approach.\medskip

Combining these results with those of the previous section, we now
have all the building blocks to discriminate $Z^\prime \to t\bar{t}$
signal events from the mixed QCD+$t\bar{t}$ background.  We can follow
two different approaches: consider the problem as signal vs background
classification or think of it as classifying events into one signal
and two background categories. For this comparison, we use the pooling
setup shown in Fig.~\ref{fig:pooling_capsnet}, as well as the
convolution setup from Fig.~\ref{fig:capsnet_conv}. Moving from one
common background label to two distinct backgrounds leads us to a
multi-class CapsNet, including the choice of estimators alluded to in
Eq.\eqref{eq:signalness}. In training the 2-class network we use a
sample with half signal and half background events, the background
further divided evenly between $t \bar{t}$ and QCD background . For
the 3-class case we use equal parts for each label. We considered
splitting the backgrounds following their respective rates, but in
this case the background sample would have been entirely dominated by
QCD. As in the previous sections the training sample consists of 300k
events combined.

\begin{figure}[t]
\centering
\includegraphics[width=0.45\textwidth]{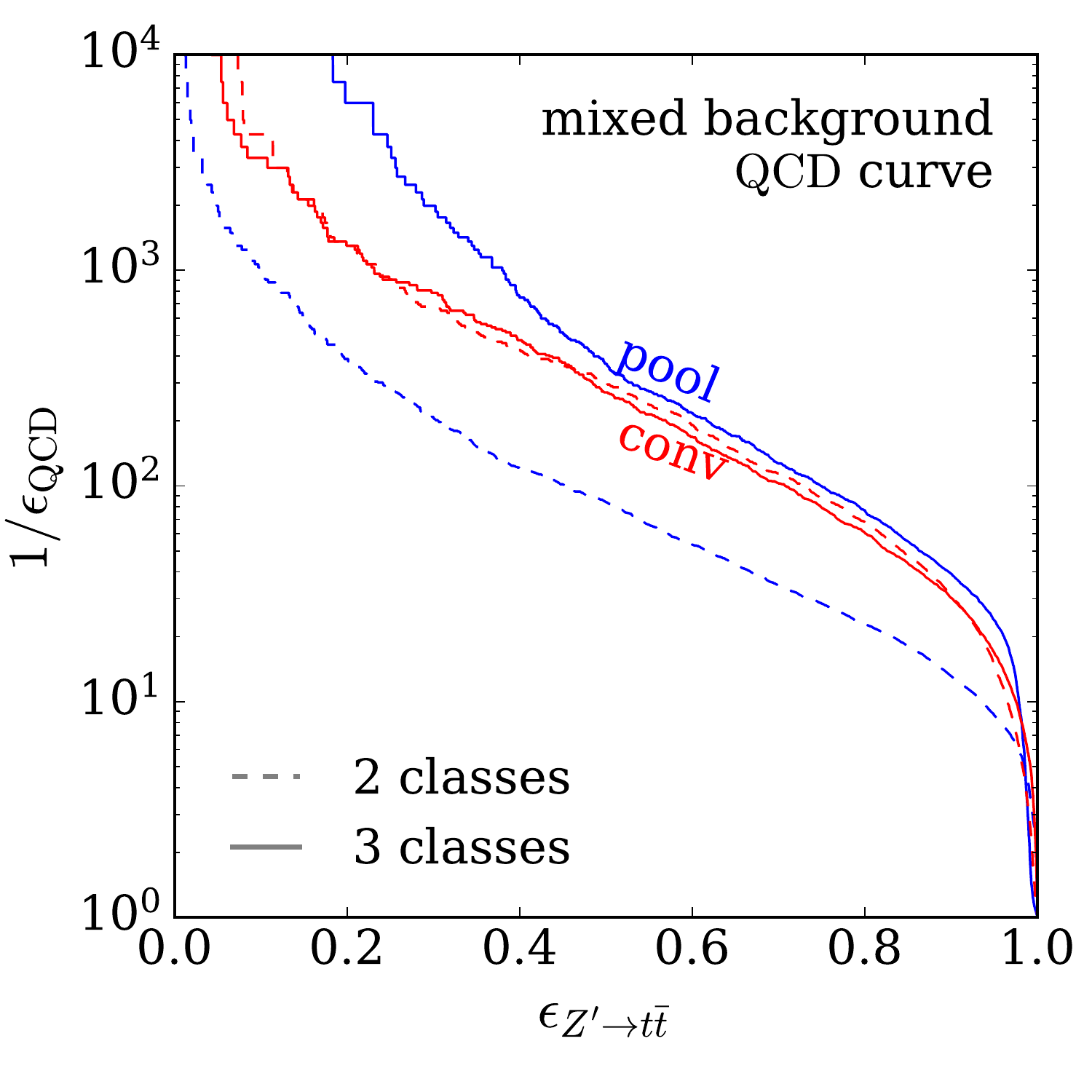}
\hspace*{0.03\textwidth}
\includegraphics[width=0.45\textwidth]{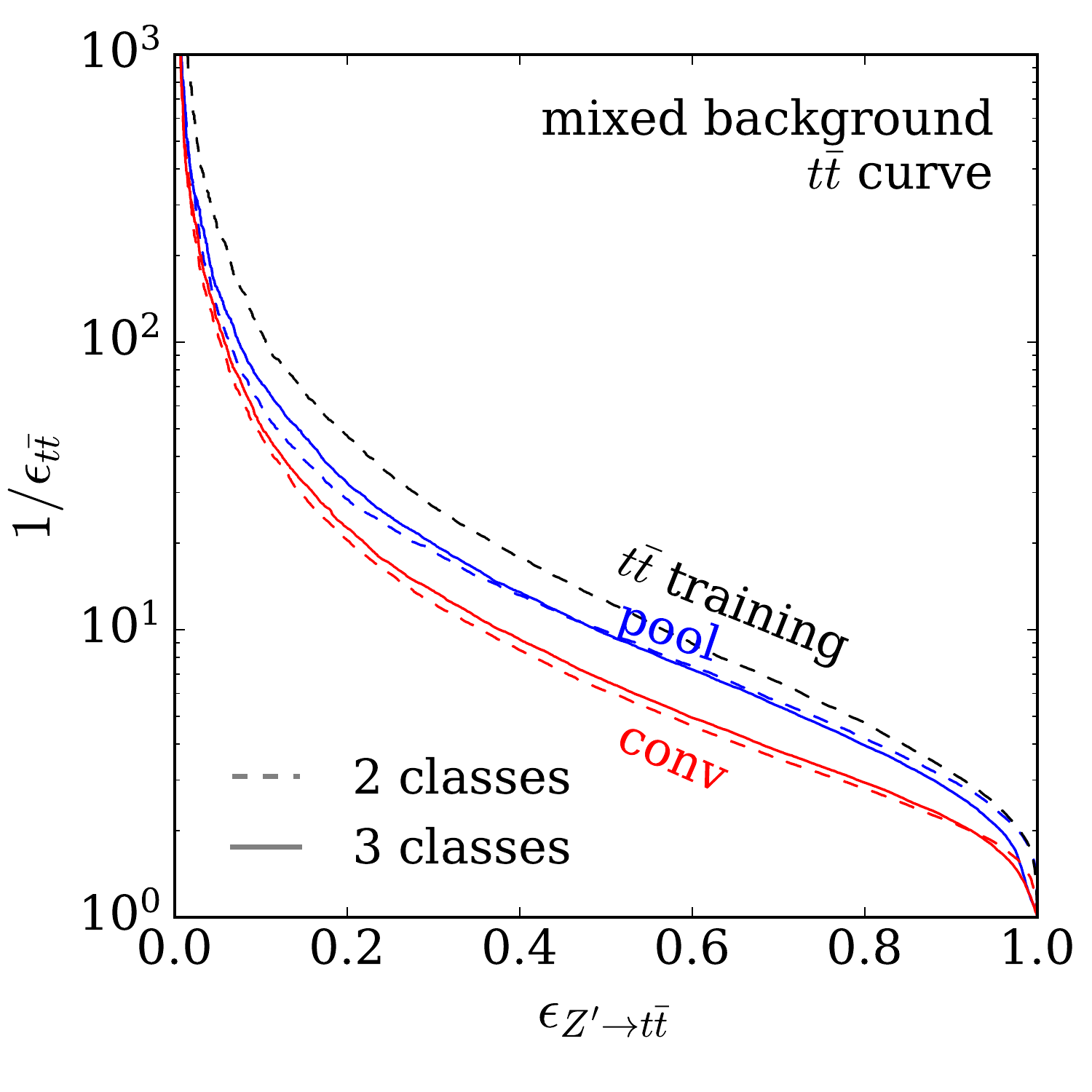}
\caption{Left: ROC curve for the QCD di-jet background for pooling and
  convolutional CapsNets, each with two or three classes. Right:
  corresponding ROC curves for the continuum $t\bar{t}$
  background. The $t\bar{t}$ training line quotes the best CapsNet
  result from Fig.~\ref{fig:vsBDT_roc}. We always train on both
  backgrounds and only separate the testing.}
\label{fig:mixedbkgroccurve}
\end{figure}

The the two panels of Fig.~\ref{fig:mixedbkgroccurve} show the
rejection of the QCD background (left) and the continuum $t\bar{t}$
background (right) by CapsNets trained on mixed background samples.
The number of classes used has different effects for the two
architectures.

For the convolutional architecture there appears to be no significant
difference between the two-class and three-class versions.  This is
because the convolutional setup is very apt at extraction subjet
features, but not good at event-level information. Given that the
combined background capsule of the two-class setup encodes subjet
features efficiently, a dedicated QCD capsule offers little
improvement. Consequently, the convolutional CapsNet performs very
weakly for the $t\bar{t}$ background rejection, and moving from two to
three classes helps very little with this structural deficit.

The situation is different for the more carefully constructed pooling
CapsNet. In QCD background rejection it very clearly benefits from the
3-class setup. The reason is that the pooling setup is designed with
event-level kinematics in mind, so when one capsule faces both
backgrounds it will focus on the event-level features and deliver a
poor QCD background rejection. In its 3-class version the pooling
setup can train a dedicated QCD capsule on the subjet features
extremely well. For the $t\bar{t}$ background rejection the pooling
CapsNet the third class leads to no improvement, because the 2-class
network already learns the event-level information.\medskip

Altogether, we find that a 3-class pooling CapsNet is best suited for
extracting the $t\bar{t}$ resonance signal from a mixed $t\bar{t}$ and
QCD background. When comparing its performance to the that from a pure
$t\bar{t}$ background in the right panel of
Fig.~\ref{fig:mixedbkgroccurve}, we still notice a slight drop in
performance for the $t\bar{t}$ background rejection. This has two
contributing factors: first, the network needs to learn 50\% more
features in going from a 2-class to a 3-class problem with the same
number of weights.  Second, the additional output class adds more
possibilities for mis-stating. The first issue can be fixed by adding
more weights up to the point where computing power becomes the
limiting factor, the second is inherent to multi-class problems.

\section{Inside capsules}
\label{sec:inside}

\begin{figure}[t]
\centering
\includegraphics[width=0.9\textwidth]{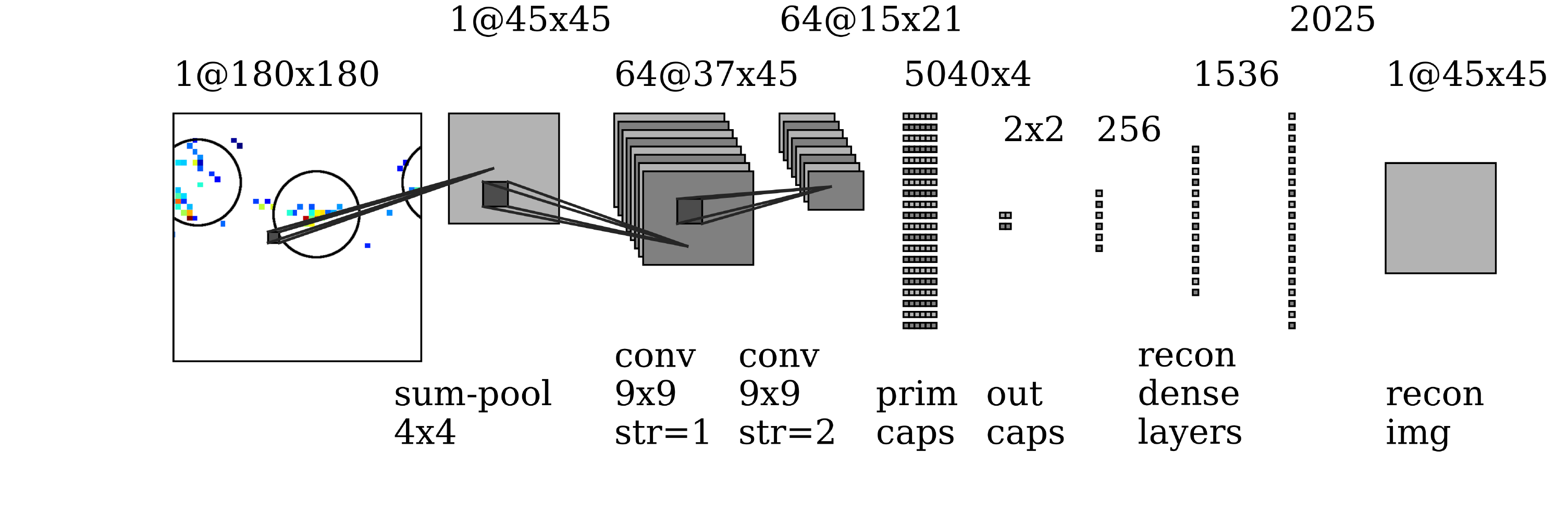}
\caption{Simplified Capsule architecture.}
\label{fig:capsnet_simp}
\end{figure}

\begin{figure}[b!]	
\centering
\includegraphics[width=0.65\textwidth]{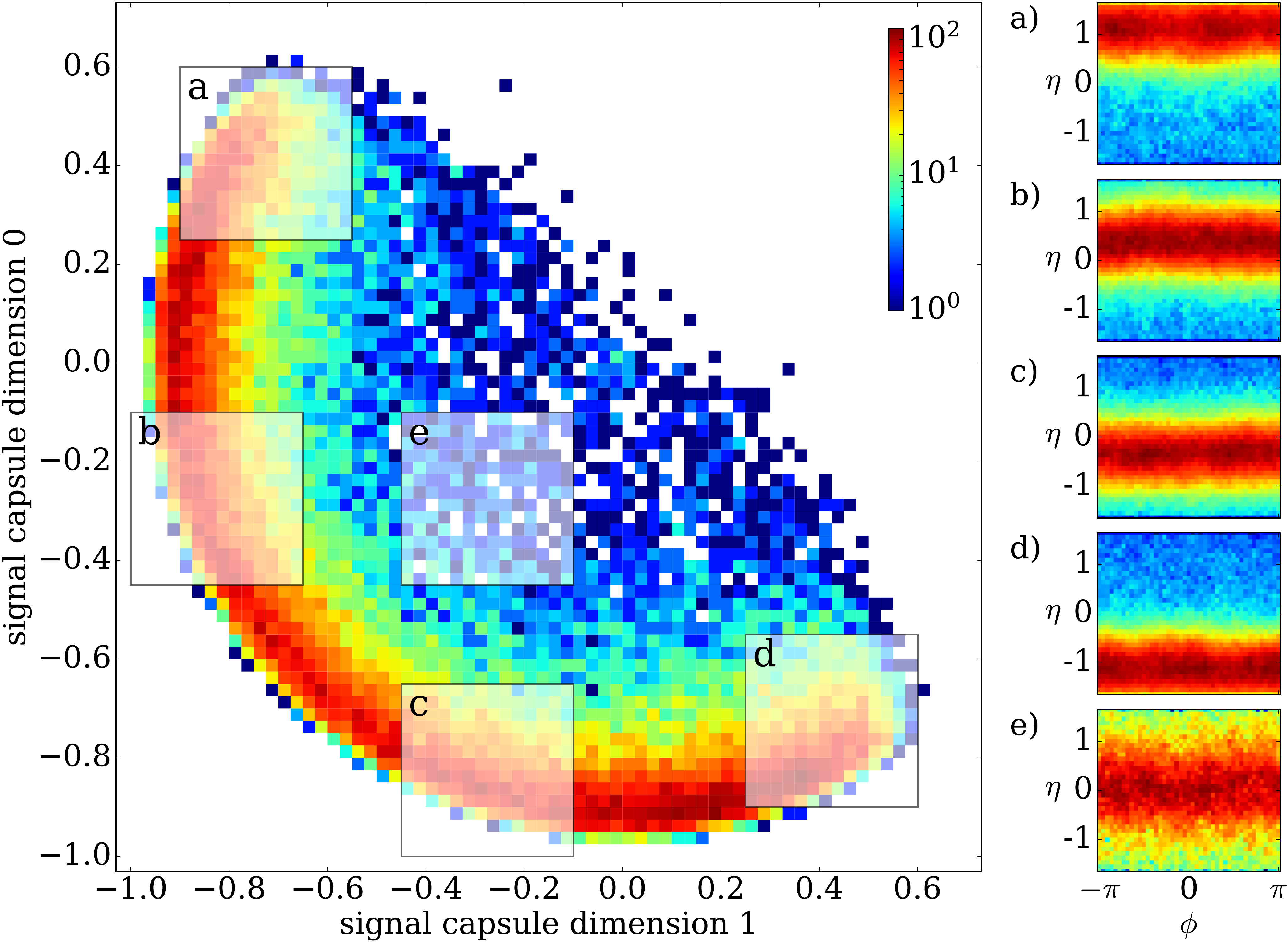}
\caption{Distribution of the two entries in the 2-dimensional signal
  capsule for signal events. Right: average event images in the
  $\eta-\phi$ plane.}
\label{fig:SigCaps}
\end{figure}

Before moving to even more complex problems, we want to understand
what the capsule vectors learn. For this visualization aspect we again
separate $Z' (\to t\bar{t})$ from QCD di-jet events. The signal and
background events then differ in event-level kinematics and in jet
substructure.  To further simplify the problem we reduce the
resolution of input images from $180 \times 180$ to $45 \times 45$
pixels using sum-pooling with a kernel size 4.  This brings us close
the size of MNIST digits of $28 \times 28$ pixels and allows us to use
an architecture very similar to the original
CapsNet~\cite{capsules1,capsules2}.

The detailed architecture is illustrated in
Fig.~\ref{fig:capsnet_simp}. The network has two output capsules $Z'$
(signal) and QCD (background) with two dimensions each. 
Inputs to the simplified model are scaled with a logarithmic function. 
As in Sec.~\ref{sec:reco}, we
train on 150,000 $Z'$ events and 150,000 QCD events with a total of
100,000 events reserved each for validation and testing. In complete
analogy to the full implementation, we use a combination of margin
loss for the capsules and MSE loss for the reconstruction network,
where for the visualization task the reconstruction network becomes
relevant. The reconstructing network achieves a classification
accuracy of 95.6\%, close to the approximately 96\% obtained by the
full network for the same problem.\medskip

\begin{figure}[t]
\centering
\includegraphics[width=0.65\textwidth]{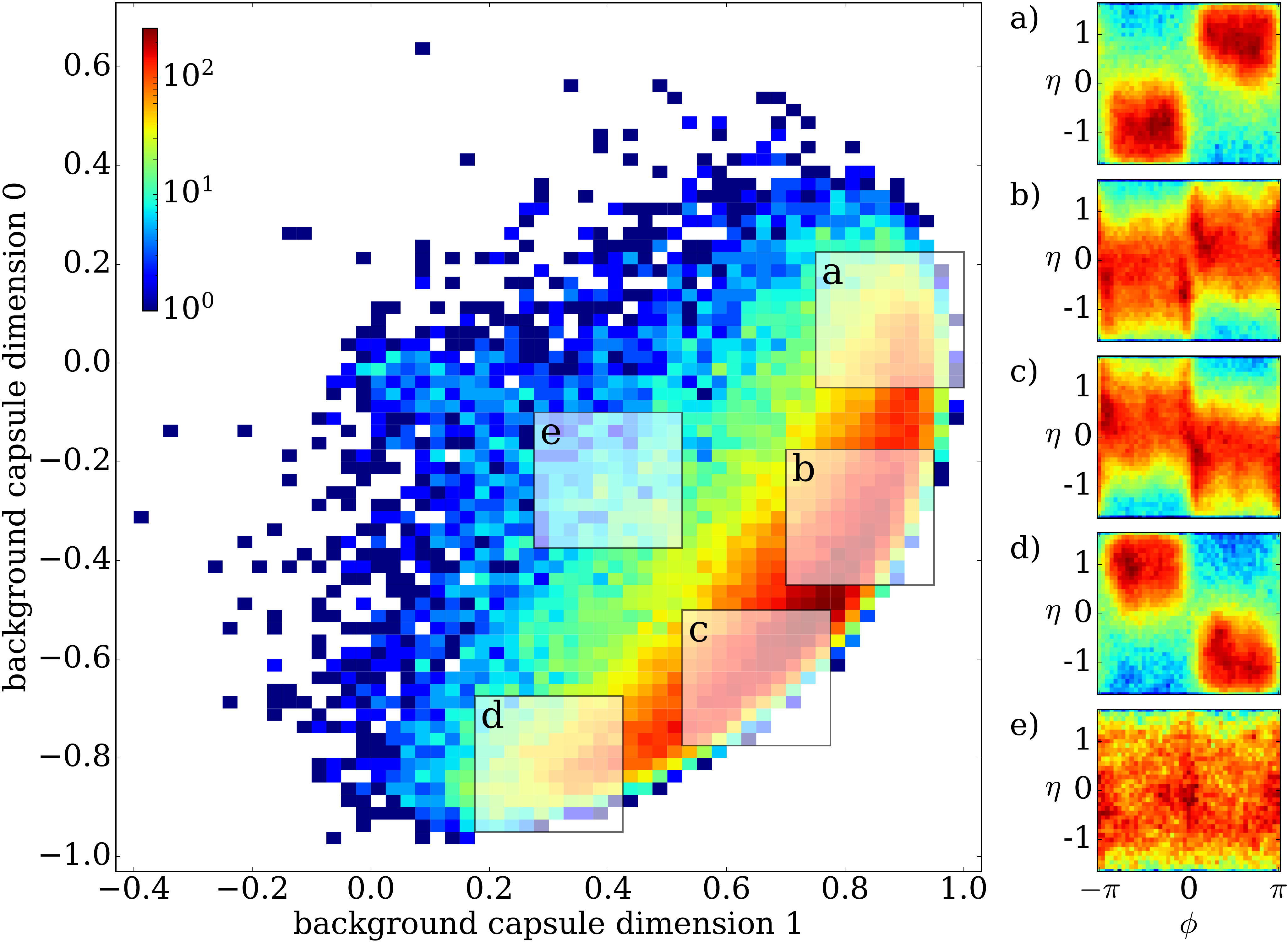}
\caption{Distribution of the two entries in the 2-dimensional
  background capsule for background events. Right: average
  event images in the $\eta-\phi$ plane.}
\label{fig:BgCaps}
\end{figure}

\begin{figure}[b!]	
\centering
\includegraphics[width=0.40\textwidth]{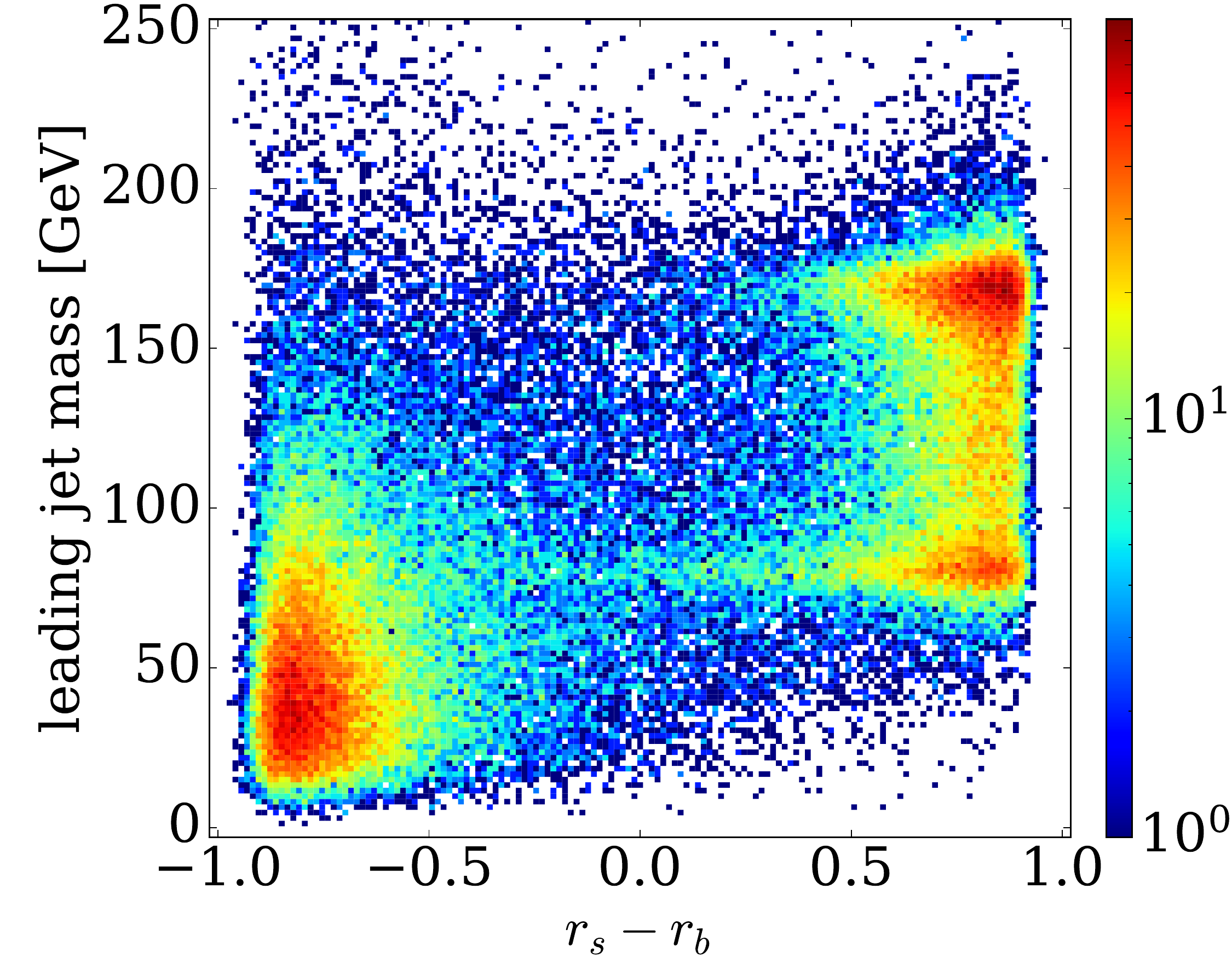}
\hspace*{0.03\textwidth}
\includegraphics[width=0.40\textwidth]{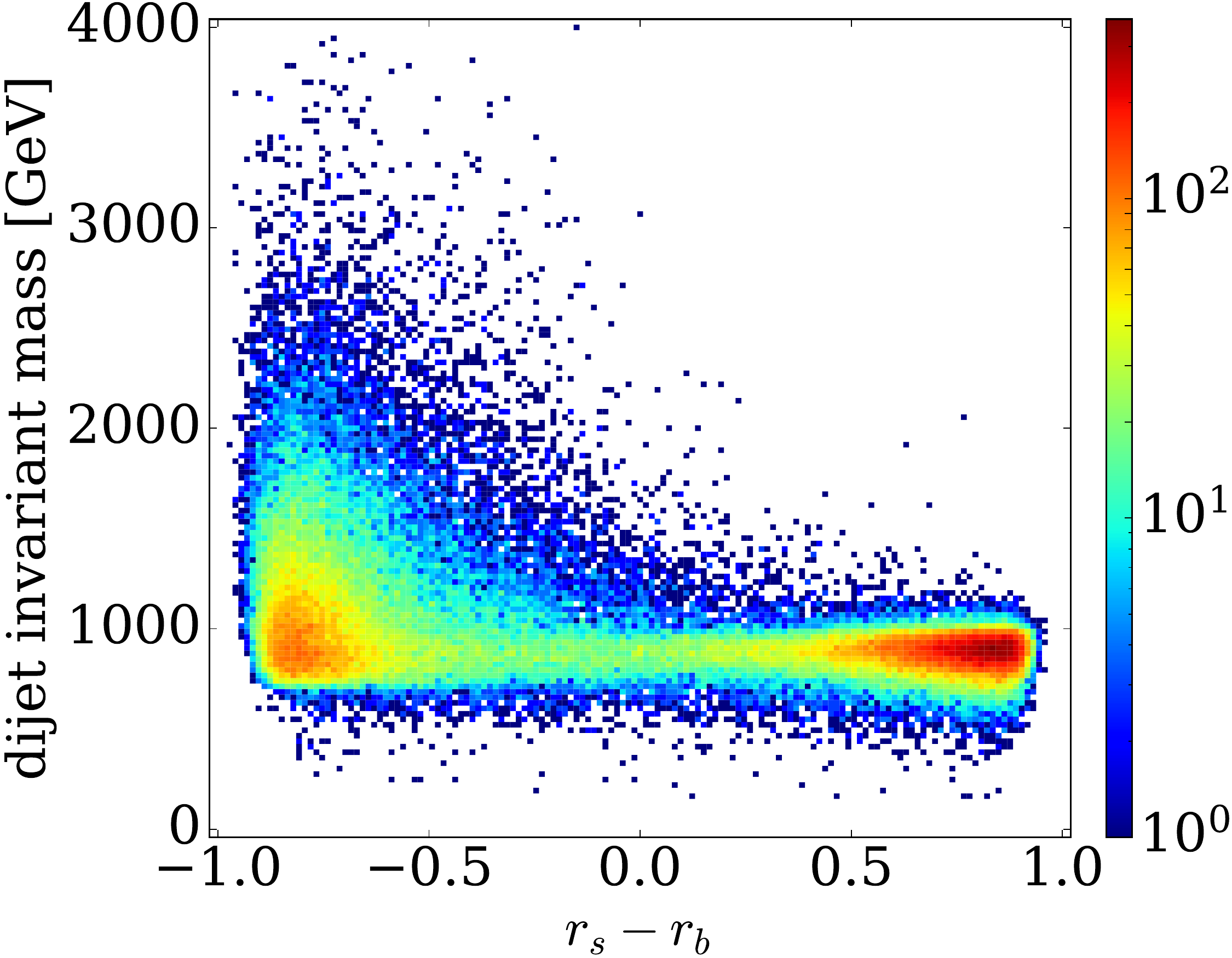} \\
\includegraphics[width=0.40\textwidth]{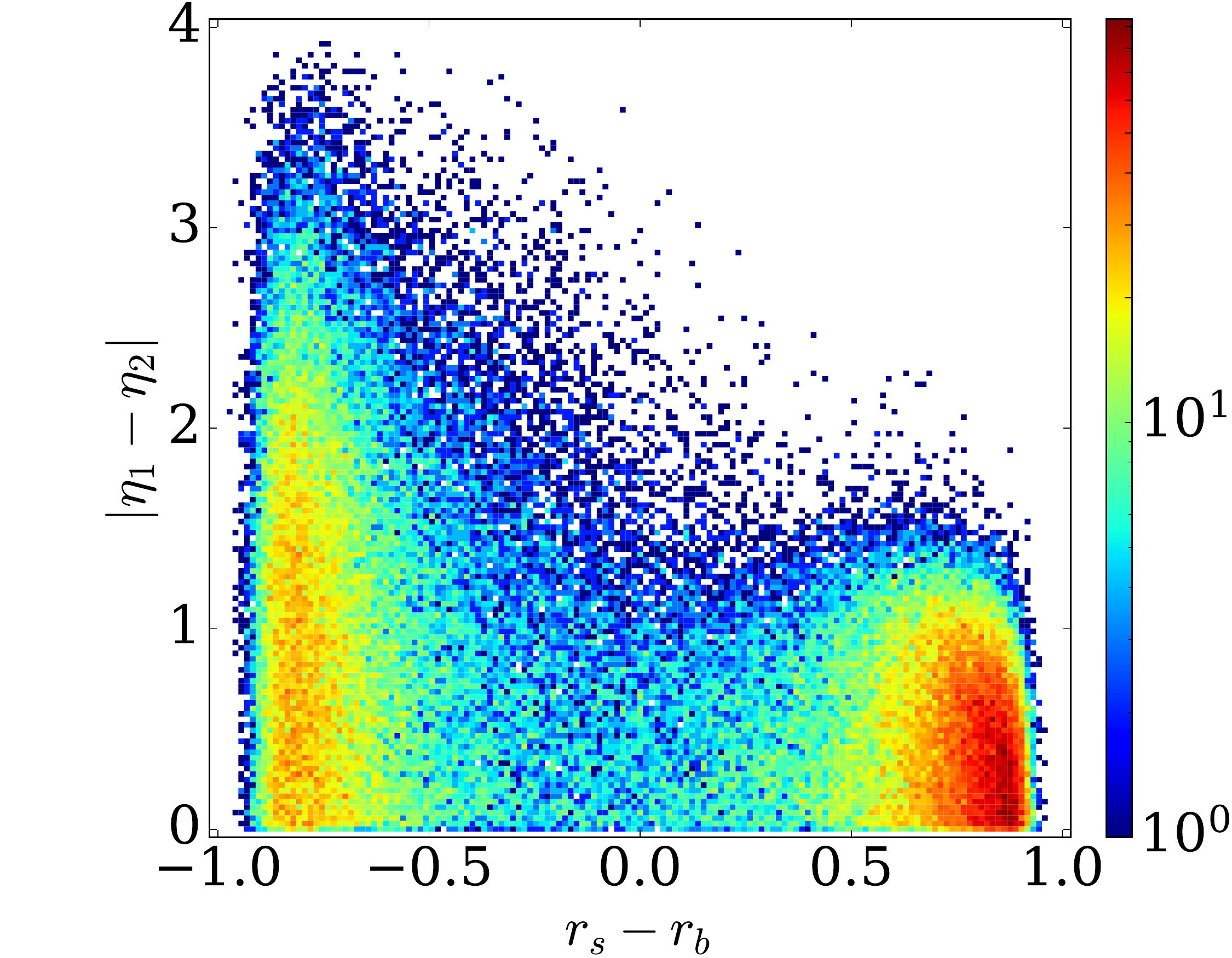}
\hspace*{0.03\textwidth}
\includegraphics[width=0.40\textwidth]{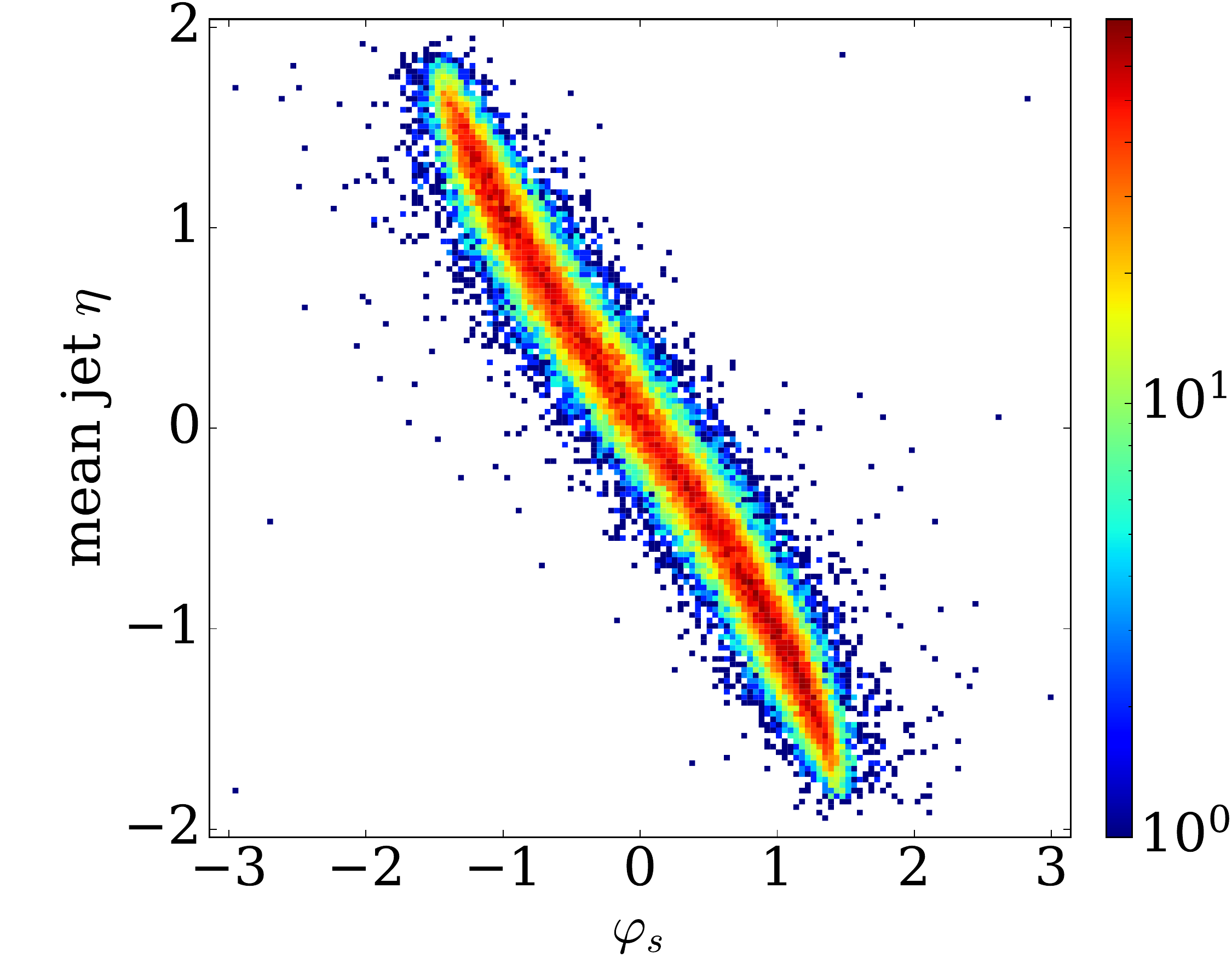}
\caption{Correlation between capsule outputs $r_S-r_B$ and the leading
  jet mass, the di-jet $m_{jj}$, and $\Delta \eta_{jj}$ for true signal
  and background events.  Finally we show the correlation of the
  signal $\varphi$ vs the the mean $\eta_j$ for true signal events.}
\label{fig:correlations}
\end{figure}

In Fig.~\ref{fig:SigCaps} we show a density plot of the two output
entries in the 2-dimensional signal capsule on true signal
events. Each event corresponds to one point in the 2D plane. If the
classification output is proportional to the length of the capsule
vector it corresponds to the distance of each point from the origin.
This explains why many events are distributed in a filled circle
segment distribution. A large fraction of events sits on the boundary
which corresponds to the most signal-like examples. The rotation of
the circle segment is not fixed a priori, each training is not
guaranteed to fill the full circle, and multiple trainings will
reproduce the same shape with different orientations.

In this 2-dimensional capsule plane we select five representative
regions indicated by semi-transparent squares. For each region we
identify the contributing events and super-impose their detector
images in the $\eta - \phi$ plane in the right panels of
Fig.~\ref{fig:SigCaps}. For the signal we observe bands for given
rapidities and smeared out in the azimuthal angle, indicating that the
network learns an event-level correlation in the two $\eta_j$ as an
identifying feature of the signal.  Figure~\ref{fig:BgCaps} gives the
same information for background capsule outputs on true background
events. We observe the same radial pattern, but the mapping on event
image reveals a very different pattern with a clear back-to-back
correlation in the rapidity vs azimuthal angle plane.\medskip

To better understand this behavior we transform the capsule outputs
into polar coordinates, where the radius $r$ encodes the
discrimination following Eq.~\eqref{eq:signalness} and $\varphi$
refers to different instantiations which do not matter for
classification. The signal-background discriminator returns
$r_S - r_B \equiv |\vec{v}^{(S)}|-|\vec{v}^{(B)}| =+1$ for maximally
signal-like events and $r_S - r_B =-1$ for maximally background-like events.
 In Fig.~\ref{fig:correlations} we first confirm
that the network identifies the large jet mass for the top signal,
where the secondary peak in the leading jet mass arises from cases
where the jet image only includes two of the three top decay jets and
learns either $m_W$ or the leading $m_{jb} \approx
m_W$~\cite{heptop2}.  Next, we see that the capsules also learn to
identify the peak in the dijet invariant mass at approximately 1~TeV
as identifying feature of signal events opposed to the kinematically
falling spectrum for background-like events. As already observed in
Figs.~\ref{fig:SigCaps} and ~\ref{fig:BgCaps}, signal jets typically
have a lower separation in $\eta$ than background jets, reflected by
the lower left panel of Fig.~\ref{fig:correlations}. Finally, we
confirm that the polar angle $\varphi_S$ for signal events perfectly
learns the absolute jet positions in $\eta$. We have checked that for
background events the jet position in $\phi$ is learned by the
corresponding background polar angle $\varphi_B$.

\section{$\mathbf{t\bar{t}H}$ production}
\label{sec:higgs}

\begin{figure}[b!]
\centering
\includegraphics[width=0.45\textwidth]{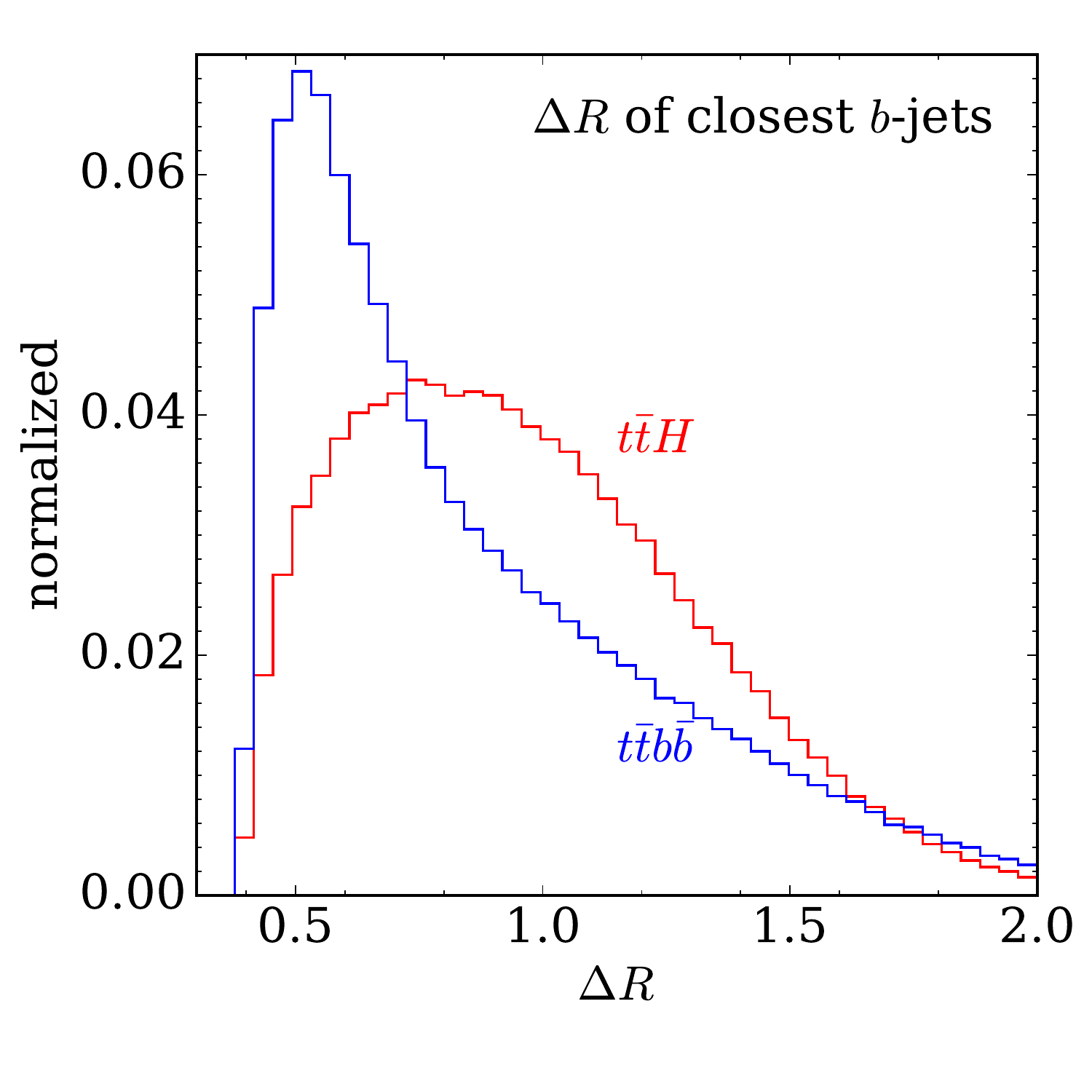}
\hspace*{0.03\textwidth}
\includegraphics[width=0.45\textwidth]{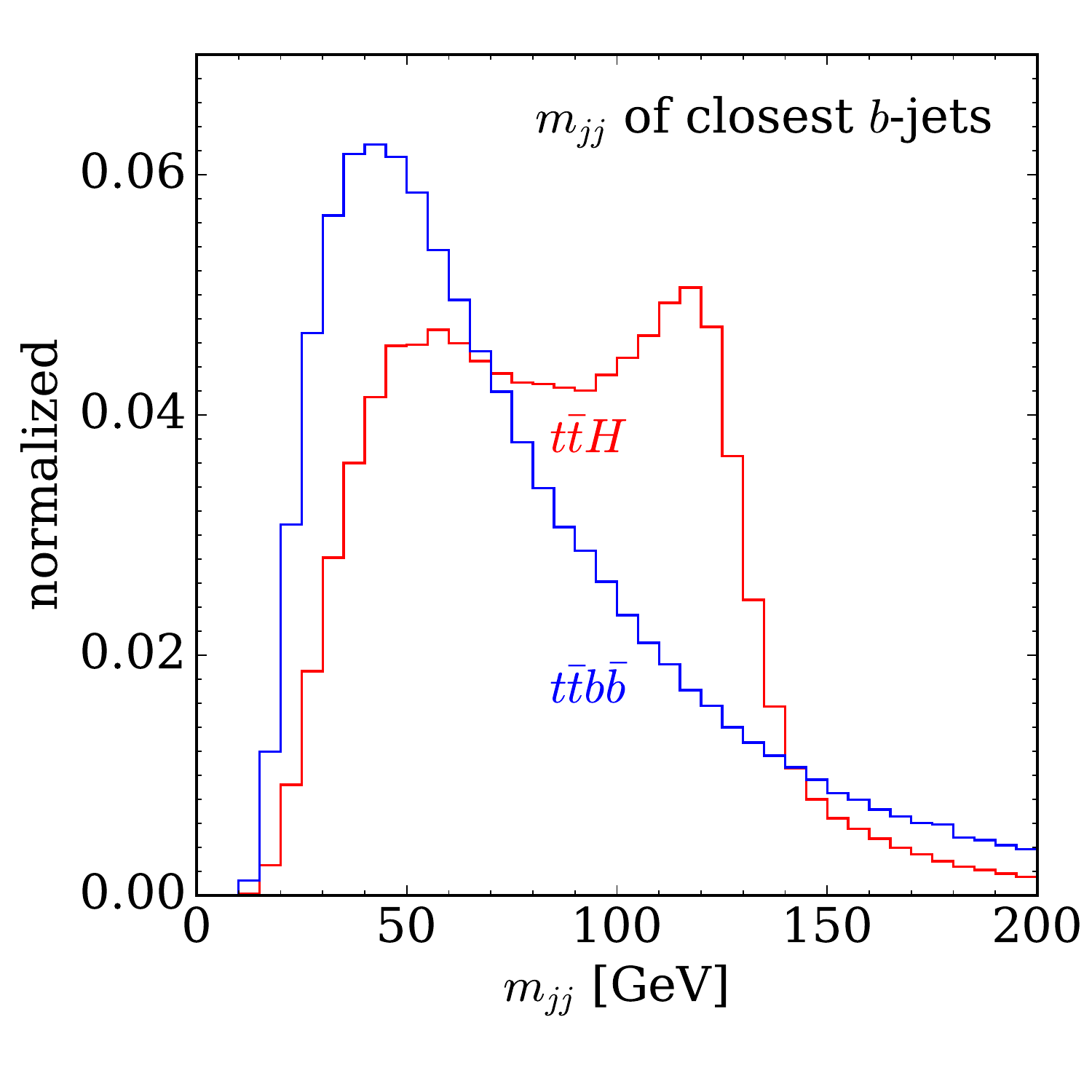}
\caption{Left: minimum $\Delta R_{bb}$ between any two
  $b$-jets. Right: invariant mass $m_{bb}$ of these two closest
  $b$-jets.}
\label{fig:ttHk_kinematics}
\end{figure}

Finally, we need to show how CapsNets can go beyond single images to
supplement calorimeter information for example with tracking
information. We illustrate this feature with one of the most complex
Standard Model signatures, namely associated top-Higgs production. It
allow us directly measure the Higgs-top interaction, which is,
arguably, the most interesting Higgs property accessible at the
LHC. The experimental challenge is that this production process comes
with a low production rate and a particularly complex final state. We
consider this signal combined with the dominant Higgs decay for a
sizeable rate and one leptonic top decay for triggering,
\begin{align}
pp \to t \bar{t} H \to t \bar{t} \; (b\bar{b}) \; .
\label{eq:ttH_signal}
\end{align}
The leading continuum background is
\begin{align}
pp \to t \bar{t} \; b \bar{b} \; ,
\label{eq:ttH_bkg}
\end{align}
making the classification an ideal task for event-level machine
learning and our a CapsNet tagger.  An event-level Lorentz boost
network has been applied to the same signal process in
Ref.~\cite{lbn}. This network is designed to construct useful
Lorentz-invariant quantities and observables from the particle
4-momenta. It is a very different approach to that considered here and
serves as an excellent benchmark for our study.\medskip

\begin{figure}[t]
\centering
\includegraphics[width=0.9\textwidth]{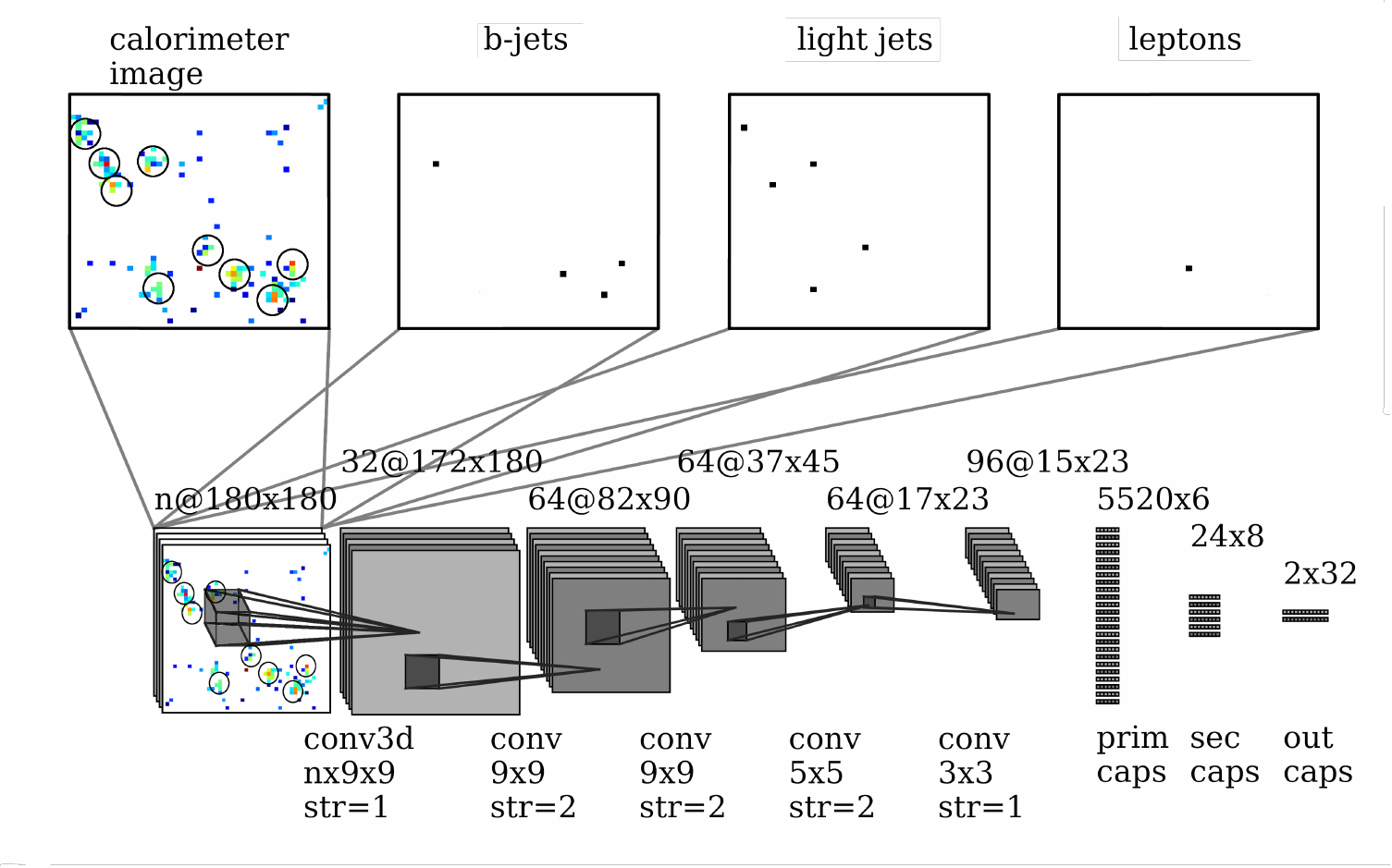}
\caption{$t \bar{t} H$ CapsNet 3D convolution architecture with
  additional jet flavor and lepton information.}
\label{fig:ttH_3d_capsnet}
\end{figure}

We generate this process with the same setup as described in
Sec.~\ref{sec:tagging}.  We enforce decays for both processes, namely
$H \to b \bar{b}$, $t \to b \ell^+ \nu_\ell$ and $\bar{t} \to \bar{b}
jj$ with $j = d,u,s,c$ and $\ell = e, \mu$. To analyze the event-level
kinematics and relate our study to standard LHC analyses we also
reconstruct jets with $R=0.4$, even though the CapsNet analyses the
calorimeter images without reference to jets. We select events with
\begin{enumerate}
\setlength\itemsep{-0.3em}
\item exactly one muon or electron with $p_{T \ell} > 5$ GeV and
  $\vert\eta_\ell\vert < 2.5$;
\item at least 6 jets (anti-$k_T$~\cite{anti_kt}, $R=0.4$) with $p_{T j}
  > 20$ GeV and $\vert\eta_{j}\vert < 2.3$; and
\item each of the 4 $b$-jets truth-matched to a $b$-parton within
  $\Delta R=0.4$.
\end{enumerate}
Because both signal and background contain four $b$-jets we do not
consider a finite $b$-tagging efficiency, as it will have no
significant impact on our conclusions. Assuming four $b$-tags we then
reconstruct the hadronic top by combining one $b$-jet with two
light-jets and minimizing $\vert m(j_b + j_1 + j_2) - m_{t} \vert$.
Because we know that the significance is dominated by the boosted
regime~\cite{heptop1}, we require the reconstructed hadronic top jet
to have $p_{T j_{t}} > 200$~GeV and $\vert m_{j_{t}} - m_{t} \vert <
30$~GeV, to avoid producing a large number of events with little
sensitivity. Our results should not depend on this slight
simplification.

For illustration, Fig.~\ref{fig:ttHk_kinematics} shows some kinematic
properties of the signal and background processes. The small
differences are difficult to exploit in a cut-based analysis.  A
reconstruction of the Higgs mass peak is at least seriously
challenging because of the $b$-combinatorics~\cite{tdr}, which is the
main motivation of a boosted analysis of this
process~\cite{heptop1,boosted}. To fully exploit these signal features
we employ our CapsNet, to show that it can both identify objects and
explore their geometric correlations.\medskip

\begin{figure}[t]
\centering
\includegraphics[width=0.45\textwidth]{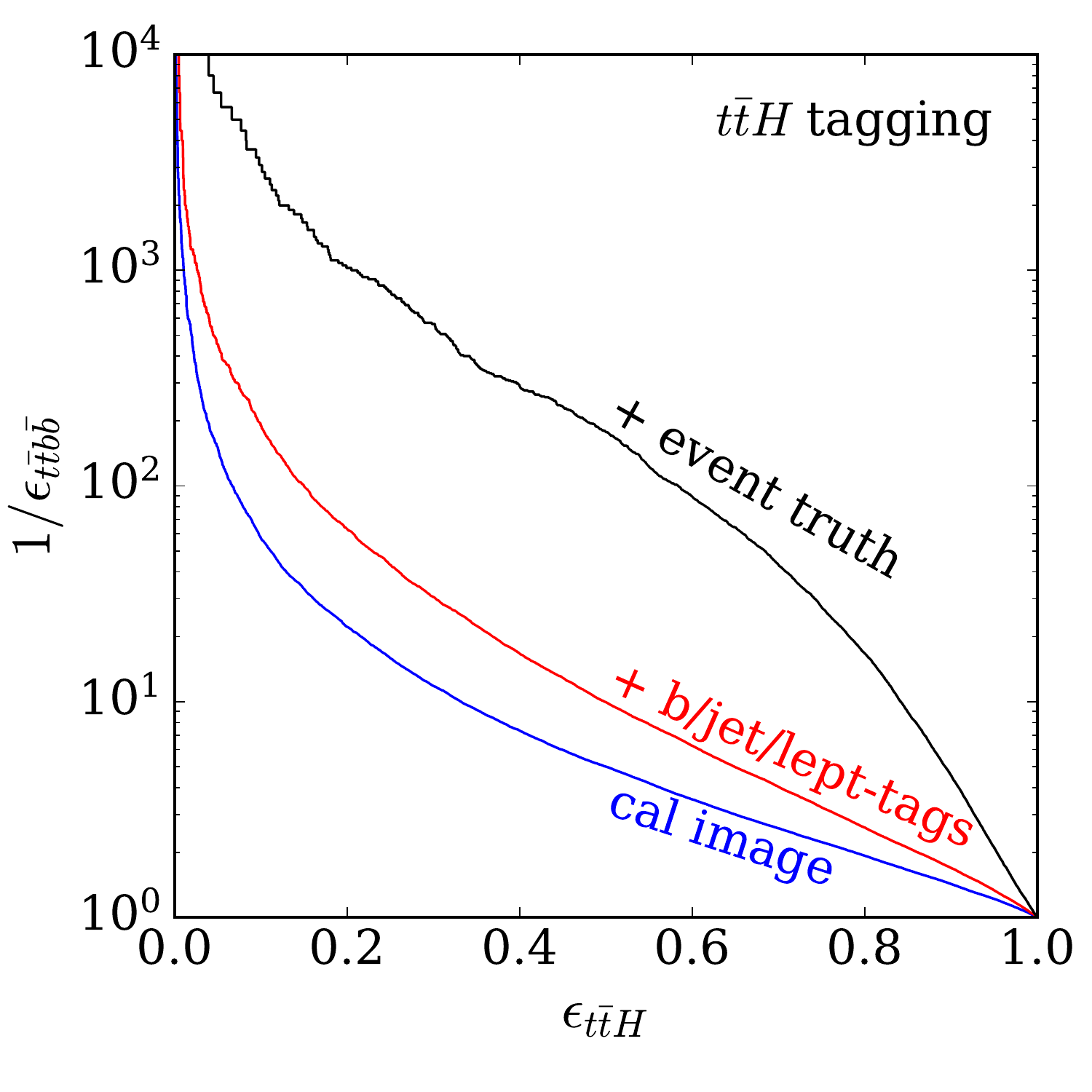}
\caption{ROC curves for $t\bar{t}H$ with calorimeter information only,
  physically accessible information and with MC truth information.}
\label{fig:ttH_roc_curve}
\end{figure}

From the previous sections we already know that we can choose a
pooling or a convolutional CapsNet to analyse the event-level
information for the complex $t\bar{t}H$ final state. We have
seen that the convolutional CapsNet well-suited for subjet studies, but we
also know that the pooling setup is superior for combining subjet and
event-level information. Because the $t\bar{t}H$ analysis does not
involve subjet information and the challenge will be to combine
overlaying images all on the event level the convolutional CapsNet
with its minimized loss of information and resolution turns out the
better-suited approach.

We illustrate this CapsNet architecture in
Fig.~\ref{fig:ttH_3d_capsnet}.  We use it to analyze our usual $(180
\times 180)$-pixel calorimeter image which pixel-wise encodes the
$E_T$.  Moreover, we want to include information from the particle
identification, such as the position of identified leptons or
$b$-tags. This information is included in the form of additional
feature maps for each physically distinct paricle class also shown in
Fig.~\ref{fig:ttH_3d_capsnet}~\cite{seeing}. We also add a feature map
with the light jet axes, which does not include any additional
information but can help the network with its sparsely filled pixels.
These feature maps are first combined through a 3-dimensional
convolution, before each of them is independently passing through the
CapsNet with its 2-dimensional convolutions. This combination of
2-dimensional and 3-dimensional convolutions allows the network to
extract information both from the individual feature maps as well as
correlations between them.\medskip

To understand what information the network is using for its signal vs
background classification, in Fig.~\ref{fig:ttH_roc_curve} we compare
three different levels of information. First, we consider calorimeter
information only, which is comparable to one of the setups in
Ref.~\cite{lbn}. For this set-up we find comparable performance to
Ref.~\cite{lbn}, with an AUC of 0.715, which is slightly above their
upper limit. The network performance is extremely poor, also
because the already challenging combinatorics of $b$-jets is worsened
by the many additional light-flavor jets.  We can improve upon this by
adding feature maps for the $b$-jets, the lepton, and potentially also
the light jet axes.  Figure~\ref{fig:ttH_roc_curve} shows how this
information improves the background rejection by a factor two to three
and gives an area under the ROC curve of AUC=0.792.  To understand
where the limitations of our analysis lies and what our network is
technically capable to handle we also add MC truth
information. Specifically, we remove the combinatorics by labelling
where each $b$-jet originates. Including this unphysical information
show that our analysis is not limited by the CapsNet performance and
gives us a ceiling in perormance of AUC=0.927.

\section{Outlook}
\label{sec:outlook}

We have demonstrated the power of capsule networks for the particle
physics task of LHC event tagging.  Their unique representation of
information makes them an ideal tool for identifying similar patterns
when the convenience of regularizing images is removed. 

While sparsely filled large number of pixels in calorimeter images are
a limiting factor for convolutional networks, CapsNets are designed
to go beyond those limitations.  They are optimized to
extract, both, low-level subjet information and event-level
kinematics at the same time. We have illustrated the capabilities of
simple CapsNets using three processes:
\begin{itemize}
\setlength\itemsep{-0.2em}
\item tagging of a $t\bar{t}$ pair using subjet information;
\item tagging and reconstructing a $Z' \to t\bar{t}$ resonance adding event-level information;
\item extracting $t\bar{t}H$ production using overlaying event-level images. 
\end{itemize}
We have employed different CapsNets, based on convolutions as well as
based on pooling, and shown that LHC signatures typically benefit from
multi-class architectures. In all of these aspects, capsules are a
natural way to go beyond standard convolutional networks. Finally, we
have shown how the CapsNet output is much more readily interpreted
than other deep-learning architectures.

\bigskip
\begin{center} \textbf{Acknowledgments} \end{center} 

First of all, we would like to thank Anja Butter for her help during
the early phase of this project. Furthermore, we acknowledge support
by the state of Baden-Württemberg through bwHPC and the German
Research Foundation (DFG) through grant no INST 39/963-1 FUGG. JT
would like to thank BMBF for funding.


\end{document}